\RequirePackage{fix-cm}
\documentclass[smallextended]{svjour3}       
\smartqed  
\usepackage{graphicx}

\journalname{Bull.\ Math.\ Bio.}
\newcommand{\D}{\mathrm{d}}
\newcommand{\rme}{\mathrm{e}}

\newcommand{\dx}{\Delta x}
\newcommand{\dt}{\Delta t}

\newcommand{\Kat}{\prescript{}{2}{K_{\mathrm{a}}}}
\newcommand{\Kaot}{\prescript{2}{1}{K_{\mathrm{a}}}}
\newcommand{\Kato}{\prescript{1}{2}{K_{\mathrm{a}}}}
\newcommand{\Kdo}{\prescript{}{1}{K_{\mathrm{d}}}}
\newcommand{\Kdt}{\prescript{}{2}{K_{\mathrm{d}}}}
\newcommand{\Kdot}{\prescript{2}{1}{K_{\mathrm{d}}}}
\newcommand{\Kdto}{\prescript{1}{2}{K_{\mathrm{d}}}}

\newcommand{\pre}{\prescript}

\newcommand{\avBo}{\overline{B}_1}

\newcommand{\avBt}{\overline{B}_2}
\newcommand{\avBs}{\overline{B}_\Sigma}

\newcommand{\Pe}{\mathrm{Pe}}

\newcommand{\Da}{\mathrm{Da}}

\newcommand{\xmin}{x_{\mathrm{min}}}
\newcommand{\xmax}{x_{\mathrm{max}}}

\newcommand{\ra}{\mathrm{a}}
\newcommand{\rd}{\mathrm{d}}

\newcommand{\fv}{\mathbf{f}}

\newcommand{\p}{\partial}
\newcommand{\wt}{\widetilde}

\usepackage{amsmath,amsfonts,amssymb,enumerate,mathtools,breqn,makecell, verbatim,epsfig, pst-grad, pst-plot, tikz, graphicx, float,caption, natbib, mathptmx}
\usepackage[english]{babel}
\usetikzlibrary{arrows.meta}

\usetikzlibrary{arrows}

\setcellgapes{3pt}
\bibpunct{(}{)}{;}{a}{,}{,}

\numberwithin{equation}{section}
\numberwithin{figure}{section}
\numberwithin{table}{section}

\begin{document}

\title{Transport Effects on Multiple-Component Reactions in Optical Biosensors
\thanks{This work was done with the support of the National Science Foundation under award number NSF-DMS 1312529. The first author was also partially supported by the National Research Council through an NRC postdoctoral fellowship.}}

\author{Ryan M. Evans        \and
        David A. Edwards 
}

\institute{R. M. Evans \at
	      Applied and Computational Mathematics Division\\
	       Information and Technology Laboratory\\ National Institute of Standards and Technology,  Gaithersburg, MD 20899, USA
             \\ \email{ryan.evans@nist.gov}      
           \and
            D. A. Edwards \at
            Department of Mathematical Sciences, University of Delaware, Newark, DE 19716, USA\\
	    \email{dedwards@udel.edu}
}

\date{Received: date / Accepted: date}

\maketitle

\begin{abstract}
Optical biosensors are often used to measure kinetic rate constants associated with chemical reactions.  Such instruments operate in the \textit{surface-volume} configuration, in which ligand molecules are convected through a fluid-filled volume over a surface to which receptors are confined.  Currently, scientists are using optical biosenors to measure the kinetic rate constants associated with DNA translesion synthesis--a process critical to DNA damage repair.  Biosensor experiments to study this process involve multiple interacting components on the sensor surface.  This multiple-component biosensor experiment is modeled with a set of nonlinear Integrodifferential Equations (IDEs).  It is shown that in physically relevant asymptotic limits these equations reduce to a much simpler set of Ordinary Differential Equations (ODEs).     To verify the validity of our ODE approximation, a numerical method for the IDE system is developed and studied.  Results from the ODE model agree with simulations of the IDE  model, rendering our ODE model useful for parameter estimation.

\end{abstract}


\section{Introduction}
\label{Introduction}

\textbf{Note}: this manuscript now appears in the Bulletin of Mathematical Biology, and may be found through the following reference:  Evans, R.M. \& Edwards, D.A. Bull Math Biol (2017) 79: 2215. https://doi.org/10.1007/s11538-017-0327-9

\begin{figure}[tbhp!]
  \centering
  \includegraphics[scale=.7]{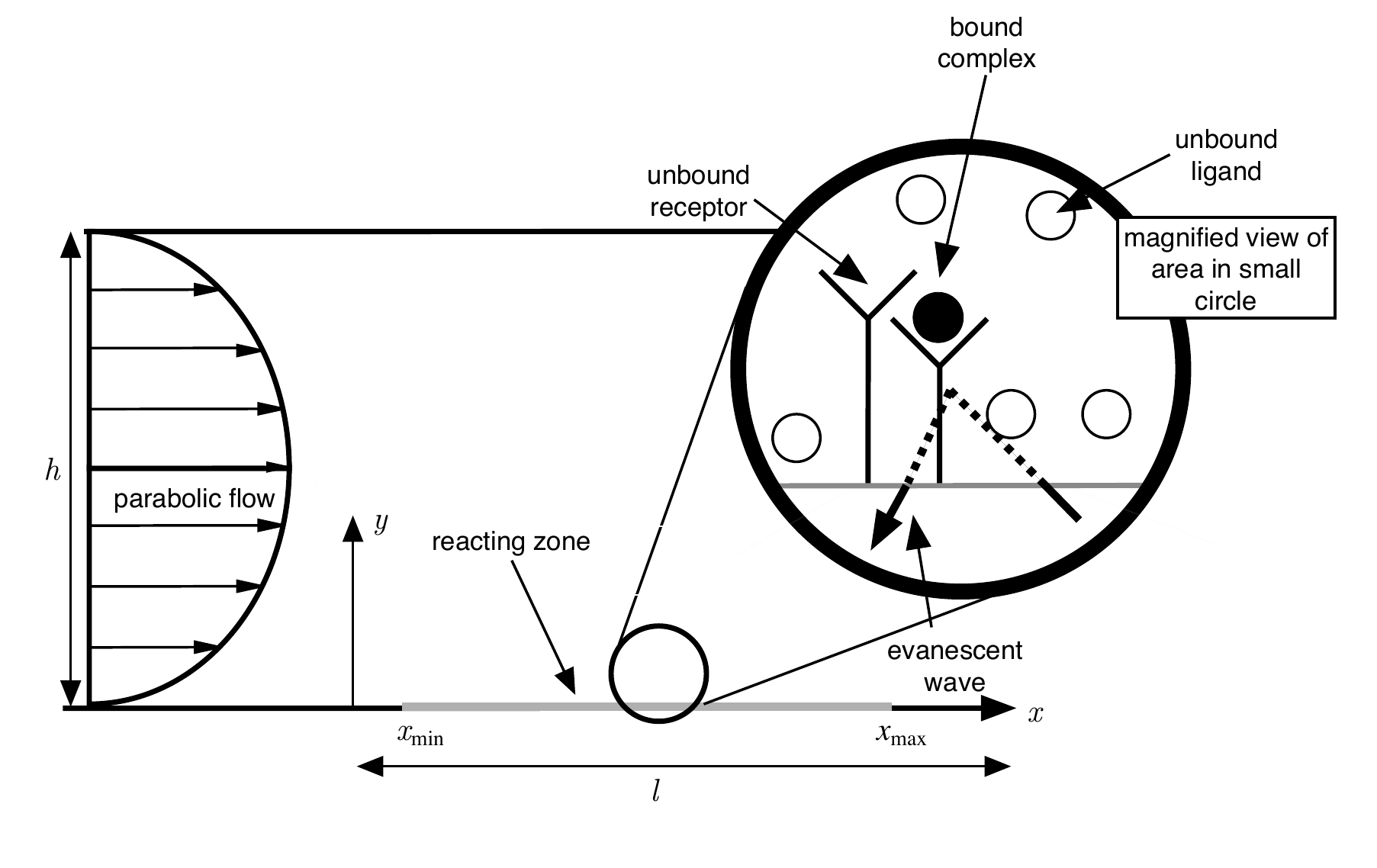}
  \caption{Cross-sectional schematic of an optical biosensor experiment.  The instrument has length $l$ and height $h$; for instrument dimensions see Appendix \ref{Appendix: Parameter Values}.  The origin corresponds to the lower left-hand corner of the instrument.  Ligand molecules are 
convected into instrument at $x=0$ in a Poiseuille flow profile, and transported to the surface to bind with receptors immobilized on reacting zone of the channel floor $[x_{\mathrm{min}},x_{\mathrm{max}}]$. }
 \label{Figure: biosensor schematic}
\end{figure}

Kinetic rate constants associated with chemical reactions are often measured using optical biosensors.  Such instruments operate in the \textit{surface-volume} configuration in which ligand molecules are convected through a fluid-filled volume, over a surface to which receptors are immobilized.  Ligand molecules are transported through the fluid onto the surface to bind with available receptor sites, creating bound ligand molecules at concentration $B(x,t)$.  Mass changes on the surface due to ligand binding are averaged over a portion of the channel floor  $[x_{\mathrm{min}},x_{\mathrm{max}}]$ to produce measurements of the form
\begin{equation}
\overline{B}(t)=\frac{1}{x_{\mathrm{max}}-x_{\mathrm{min}}}\int_{x_{\mathrm{min}}}^{x_{\mathrm{max}}}B(x,t)\ \mathrm{d}x.\label{average concentration}
\end{equation}
See Figure \ref{Figure: biosensor schematic} for a schematic of one such biosensor experiment.

Measuring kinetic rate constants with optical biosensors requires an accurate model of this process, and models have been successfully proposed and progessively refined throughout the years:  \citep{edwards1999estimating, edwards2000biochemical, edwards2001effect,  edwards2006convection, edwards2011transport, edwards1999transport,  lebedev2006convection, zumbrum2014multiple, zumbrum2015conformal}.  Although such models are typically limited to reactions involving only a single molecule or a single step, chemists are currently using biosensor technology to measure rate constants associated with reactions involving multiple interacting components.  In particular, chemists are now using biosensor experiments to elucidate how cells cope with DNA damage. Harmful DNA lesions can impair a cell's ability to replicate DNA, and its ability to survive.  One way a cell may respond to a DNA lesion is through DNA translesion synthesis \citep{friedberg2005suffering,  lehmann2007translesion, plosky2004switching}.  For a description of this process we refer the interested reader to the references included herein; however, for our purposes it is sufficient to know that DNA translesion synthesis involves three interacting components: a Proliferating Cell Nuclear Antigen (PCNA) molecule, polymerase $\delta$, and polymerase $\eta$.  Moreover, in order for a successful DNA translesion synthesis event to occur  polymerase $\eta$ must bind with the PCNA molecule.  A central question surrounding DNA translesion synthesis is whether the polymerase $\eta$ and PCNA complex forms through direct binding, or through a  catalysis-type \textit{ligand switching process} \citep{zhuang2008regulation}.

\begin{figure}[tbhp!]

\center\begin{tikzpicture}[scale=.9 , every node/.style={scale=.9}]

\node[label=below:{\makecell[l]{$EL_1$\\ $(B_1)$}}] (1) at (-.75,0) {};
\node[label=below:{\makecell[l]{$EL_1L_2$\\ $(B_{12})$}}] (2) at (2.25,0) {} ;
\node[label=below:{\makecell[l]{\textcolor[rgb]{0,0,1}{$EL_2$}\\ \textcolor[rgb]{0,0,1}{$(B_2)$}}}] (3) at (5.25,0) {};
\node[] (4) at (2.25,0) {};
\node[] (5) at (-.75,-.05) {};
\node[] (6) at (2.25,.1) {};
\node[] (7) at (-.75,.1) {};
\node[] (8) at (5.25,.1) {};
\node[] (9) at (2.25,-.05) {};
\node[] (10) at (-1.65,3.15) {};
\node[] (11) at (-.75,1.8) {};
\node[] (12) at (-.9,1.9) {};
\node[] (13) at (-1.75,3.05) {};
\node[] (14) at (5.25,1.8) {};
\node[] (15) at (6.15,3.15) {};
\node[] (16) at (5.4,1.9) {};
\node[] (17) at (6.26,3.05) {};

\draw (-3.5,0) -- (8,0);

\draw (-.75,0)-- (-.75,1);
\draw(-.75,1)--(-1.75,2);
\draw(-.75,1)--(.25,2);
\draw[black,fill=gray!10] (-.75,1.5) circle [radius=.35] node {$L_1$};
\draw[black,fill=gray!10] (-1.9,3.3) circle [radius=.35] node {$L_1$};

\draw (2.25,0)--(2.25,1);
\draw (2.25,1)--(1.25,2);
\draw (2.25,1)--(3.25,2);
\draw[black,fill=gray!10] (2.25,1.5) circle [radius=.35] node {$L_1$};
\draw[black,fill=blue!10] (2.72,2) circle [radius=.35] node {$L_2$};

\draw(5.25,0)--(5.25,1);
\draw(5.25,1)--(4.25,2);
\draw(5.25,1)--(6.25,2);
\draw[black,fill=blue!10] (5.25,1.5) circle [radius=.35] node {$L_2$};
\draw[black,fill=blue!10] (6.4,3.3) circle [radius=.35] node {$L_2$};

\path[-{Latex[left]}] 
(7) edge [bend left] node[above]{\textcolor[rgb]{0,.75,0}{${}_2^1k_{\mathrm{a}}$}} (6)  
(4) edge [bend right, dashed] node[below]{\textcolor[rgb]{1,0,0}{${}_2^1k_{\mathrm{d}}$}} (5) 
(6) edge  [bend left,dashed] node[above]{\textcolor[rgb]{1,0,0}{${}_1^2k_{\mathrm{d}}$}} (8)
(3) edge  [bend right] node[below]{\textcolor[rgb]{0,.75,0}{${}_1^2k_{\mathrm{a}}$}} (9);

\path[-{Latex[left]}] (10) edge[bend left] node[above] {\hspace{2em}\textcolor[rgb]{0,.75,0}{${}_1 k_{\mathrm{a}}$}} (11);

\path[-{Latex[left]}] (12) edge[bend right, dashed] node[below left] {\hspace{2em}\textcolor[rgb]{1,0,0}{${}_1 k_{\mathrm{a}}$}} (13);

\path[-{Latex[left]}] (14) edge[bend left, dashed] node[left] {\hspace{2em}\textcolor[rgb]{1,0,0}{${}_2 k_{\mathrm{d}}$}} (15);


\end{tikzpicture}

\caption{Schematic of the ligand switching process.  First $L_1$ binds with an available receptor $E$; next, $L_2$ associates with $EL_1$ to create the product $EL_1L_2$; then, $L_1$ dissociates from the complex $EL_1L_2$ to leave $EL_2$; finally, $L_2$ dissociates from $EL_2$.  Below each of the species $EL_1$, $EL_1L_2$, and $EL_2$, we have listed their corresponding concentrations $B_1$, $B_{12}$, and $B_2$.}
\label{Figure: ligand switching}
\end{figure}
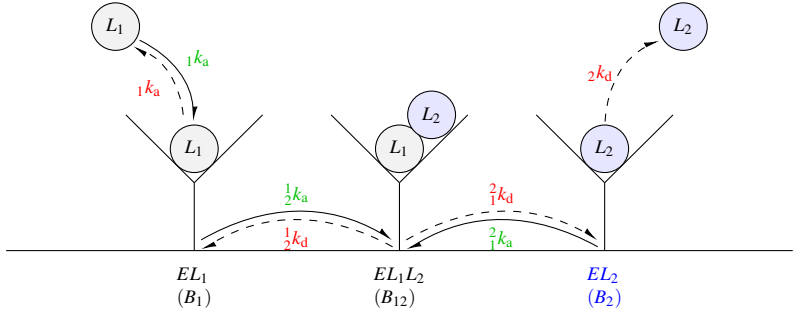

The former scenario is depicted in Figure \ref{Figure: direct binding}, where we have shown polymerase $\eta$ directly binding with a PCNA molecule, 
\textit{i.e.} the reaction:
\begin{subequations}
\begin{align}
&\mathrm{P}_1: E+L_2\xrightleftharpoons[{}_2k_{\mathrm{d}}]{{}_2
k_{\mathrm{a}}}  E L_2.\label{rxn 1}
\end{align}
Here, we have denoted the PCNA molecule and polymerase $\eta$ as $E$ and $L_2$ respectively.  Additionally, ${}_2k_{\ra}$ denotes the rate at which $L_2$ binds with an empty receptor $E$, and ${}_2k_{\rd}$ denotes the rate at which $L_2$ dissociates from a receptor $E$.  We will refer to this as pathway one, or simply $\mathrm{P}_1$ as in (\ref{rxn 1}).

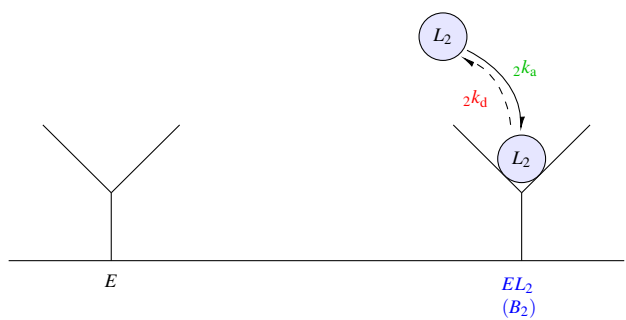
\begin{figure}[tbhp!]
\center
\begin{tikzpicture}[scale=.9 , every node/.style={scale=.9}]

\node[label=below:{$E$}]  (1) at (-3,0) {};
\node[text width=1em,align=center,label=below:{\makecell[l]{\textcolor[rgb]{0,0,1}{$EL_2$}\\ \textcolor[rgb]{0,0,1}{ $(B_2)$ }}}] (2) at (3,0) {} ;
\node (3) at (2.1,3.15) {} ;
\node (4) at (3,1.8){};
\node (5) at (2,3.05) {} ;
\node (6) at (2.85,1.9){};

\draw (-4.5,0) -- (4.5,0);

\draw (-3,0)--(-3,1);
\draw (-3,1)--(-4,2);
\draw (-3,1)--(-2,2);

\draw (3,0)-- (3,1);
\draw(3,1)--(2,2);
\draw(3,1)--(4,2);
\draw[black,fill=blue!10] (3,1.5) circle [radius=.35] node {$L_2$};
\draw[black,fill=blue!10] (1.85,3.3) circle [radius=.35] node {$L_2$};

\path[-{Latex[left]}] 
(3) edge [bend left] node[above] {\hspace{2em}\textcolor[rgb]{0,.75,0}{${}_2 k_{\mathrm{a}}$}} (4)
(6) edge [bend right, dashed] node[below left] {\hspace{2em}\textcolor[rgb]{1,0,0}{${}_2 k_{\mathrm{d}}$} } (5);

\end{tikzpicture}

\caption{Left: an empty receptor $E$.  Right: direct binding of $L_2$  with an empty receptor $E$.  This creates the product $EL_2$.  The function $B_2$ is the concentration of $EL_2$.}

\label{Figure: direct binding}
\end{figure}

The catalysis-type ligand switching process is depticted in Figure \ref{Figure: ligand switching} and stated precisely as:
\begin{align}
&\mathrm{P}_2: E+L_1 \xrightleftharpoons[{}_1k_{\mathrm{d}}]{{}_1k_{\mathrm{a}}} E L_1,\ 
EL_1+L_2 \xrightleftharpoons[{}^1_2k_{\mathrm{d}}]{{}^1_2k_{\mathrm{a}}} E L_1 L_2 \xrightleftharpoons[{}^2_1k_{\mathrm{a}}]{{}^2_1k_{\mathrm{d}}} E L_2+L_1 ,\ EL_2\xrightharpoonup{{}_2k_{\mathrm{d}}}E+L_2.\label{rxn 2}
\end{align}\label{rxns}\end{subequations}
In (\ref{rxn 2}) and Figure \ref{Figure: ligand switching} we have denoted polymerase $\delta$ as $L_1$.  This process is summarized as follows: first $L_1$ binds with an available receptor $E$; next $L_2$ associates with $EL_1$ to create the product $EL_1L_2$; then $L_1$ dissociates from $EL_1L_2$, leaving $EL_2$; finally, $L_2$ dissociates from $EL_2$.  Furthermore, in (\ref{rxn 2}) and Figure 
\ref{Figure: ligand switching} the rate constants ${}_1k_{\ra}$ and ${}_1k_{\rd}$ denote the rates at which $L_1$ binds and unbinds with a receptor $E$,  ${}^j_ik_{\ra}$ denotes the rate at which ligand $L_i$ binds with the product $EL_j$, and ${}^j_ik_{\rd}$ denotes the rate at which $L_i$ dissociates from the product $EL_1L_2$. In the latter two expressions the indices $i$ and $j$ can equal one or two. We shall refer to this pathway two, or simply $\mathrm{P}_2$ as in (\ref{rxn 2}).

Though Zhuang \textit{et al}. provided indirect evidence of the ligand switch in \citep{zhuang2008regulation},  a direct demonstration of this process has not been possible with conventional techniques such as fluorescence microscopy, since such techniques introduce the possibility of modifying protein activity.  Hence, scientists are using label-free optical biosensors  to measure the rate constants in (\ref{rxns}).  By measuring the rate constants in (\ref{rxns}), one could determine whether \textit{EL${}_1$L${}_2$}  forms through direct binding, or the catalysis-type ligand switching process.  We note that the latter manifests itself mathematically with ${}_2k_{\mathrm{a}}=0$, while the former with ${}_1^2k_{\mathrm{a}}= {}_1^2k_{\mathrm{d}}= {}_2^1k_{\mathrm{a}}= {}_2^1k_{\mathrm{d}}=0$.

However, the presence of multiple intereacting components on the sensor surface complicates parameter estimation.  In the present scenario there are three species $EL_1$, $EL_1L_2$, and $EL_2$ at concentrations $B_1(x,t)$, $B_{12}(x,t)$, and $B_2(x,t)$, and since optical biosensors typically measure only mass changes at the surface, lumped measurements of the form
\begin{equation}
\mathcal{S}(t)=s_1\overline{B}_1(t)+\left(s_1+s_2\right)\overline{B}_{12}(t)+s_{2}\overline{B}_2(t) \label{dim senso}
\end{equation}
are produced.  In (\ref{dim senso})
\begin{equation}
\overline{B}_{i}(t)=\frac{1}{x_{\mathrm{max}}-x_{\mathrm{min}}}\int_{x_{\mathrm{min}}}^{x_{\mathrm{max}}}B_i(x,t)\ \mathrm{d}\mathrm{x} \label{Bi avg}
\end{equation}
denotes the average reacting species concentration, for $i=1,\ 12,\ 2$, and $s_i$ denotes the molecular weight of $L_i$. The lumped signal (\ref{dim senso}) raises uniqueness concerns, since more than one set of rate constants may possibly correspond to the same signal (\ref{dim senso}).  Fortunately, through varying the uniform in-flow concentrations of the ligands,  $C_1(0,y,t)=C_{1,\mathrm{u}}$ 
and $C_2(0,y,t)=C_{2,\mathrm{u}}$, one may resolve this ill-posedness in certain physically relevant scenarios (Evans, R. M. and Edwards, D. A. and Li, W., submitted).  This approach to identifying the correct set of rate constants in the presence of ambiguous data is related to the ``global analysis'' technique in biological literature \citep{karlsson1997experimental, morton1995interpreting}.

The presence of multiple interacting species and the lumped signal (\ref{dim senso}) complicate parameter estimation even for systems accurately descibed by the well-stirred kinetics approximation.  However in \citep{edwards1999estimating}, Edwards has shown that transport dynamics affect ligand binding in a thin boundary layer near the sensor surface.  Hence, we begin in Section \ref{Section: Governing Equations} by summarizing the relevant boundary layer equations, which take the form of a set of nonlinear Integrodifferential Equations (IDEs).   In Section \ref{Section: Effective Rate Constant Approximation}, it is shown that in experimentally relevant asymptotic limits our IDE model reduces to a much simpler set of Ordinary Differential Equations (ODEs) which can be used for parameter estimation.  To verify the accuracy of our ODE approximation, a numerical method is developed in Subsection \ref{Subsection: semi-implicit finite difference algorithm}.  Convergence properties are examined in Subsection \ref{Subsection: convergence study}, and in Section \ref{Section: Effective Rate Constant Approximation Verification} the accuracy of our ODE approximation is verified by comparing results of our ODE model with results from our numerical method described in Section \ref{Section: Numerics}.  Conclusions and plans for future work are discussed in Section \ref{Section: Conclusions}.

\section{Governing Equations}\label{Section: Governing Equations}

For our purposes, biosensor experiments are partitioned into two phases: an injection phase, and a wash phase.  During the injection phase $L_1$ and $L_2$ are injected into the biosensor via a buffer fluid at the uniform concentrations $C_1(x,y,0)=C_{1,\mathrm{u}}$ and $C_2(x,y,t)=C_{2,\mathrm{u}}$.  Injection continues until the signal (\ref{dim senso}) reaches a steady-state, at which point the biosensor is washed with the buffer fluid--this is the wash phase of the experiment.  Only pure buffer is flowing through biosensor during the wash phase, not buffer containing ligand molecules.  This causes all bound ligand molecules at the surface to dissociate and flow out of the biosensor, thereby preparing the device for another experiment.  We first summarize the governing equations for the injection phase.

\subsection{Injection Phase}

To present our governing equations we introduce the dimensionless variables:
\begin{align}
\begin{aligned}
 &\wt{x}=\frac{x}{L},\ \wt{y}=\frac{y}{H},\ \wt{t}={}_1k_{\mathrm{a}}C_{1,\mathrm{u}}t,\ \wt{B}_i(x,t)=\frac{B_i(x,t)}{R_{\mathrm{T}}},\ \wt{C}_i(x,y,t)=\frac{C_i(x,y,t)}{C_{i,\mathrm{u}}},\\
 &{}_i^j\wt{K}_{\ra}=\frac{C_{i,\mathrm{u}}\cdot {}_i^jk_{\ra}}{C_{1,\mathrm{u}}\cdot {}_1k_{\ra}},\ {}_i^j\wt{K}_{\rd}=\frac{k_{\rd}}{C_{1,\mathrm{u}}\cdot {}_1k_{\ra}},\ \wt{F}_{\mathrm{r}}=\wt{C}_{\mathrm{r}}\wt{D}_{\mathrm{r}},\ \wt{C}_{\mathrm{r}}=\frac{C_{1,\mathrm{u}}}{C_{2,\mathrm{u}}},\ \wt{D}_{\mathrm{r}}=\frac{D_1}{D_2}.
 \end{aligned} 
 \label{scalings}
\end{align} 
We have scaled the spatial variables with the instrument's dimensions, time with the association rate of $L_1$ onto an empty receptor, the bound ligand concentrations $B_i$ with the initial free receptor concentration, and the unbound ligand concentrations with their respective uniform inflow concentrations.  The rate constants ${}_i^j\wt{K}_{\mathrm{a}}$ and ${}_i^j\wt{K}_{\mathrm{d}}$ are the dimensionless analogs of ${}_i^jk_{\ra}$ and ${}_i^jk_{\rd}$.  In the latter expressions the index $i=1,\ 2$, whereas $j=1,\ 2,$ or can be blank.  Furthermore, $\wt{F}_{\mathrm{r}}$ measures the diffusion strength of each reacting species, as characterized by the product of the input concentrations and the diffusion coefficients.  \textit{Henceforth, we shall drop the tildes on our dimensionless variables for simplicity.}  In particular,
we denote the dimensionless sensogram reading as
\begin{equation}
S(t)=\frac{\mathcal{S}(t)}{R_{\mathrm{T}}\cdot s_1}=\overline{B}_1(t)+\left(1+\frac{s_2}{s_1} \right)\overline{B}_{12}(t)+\frac{s_{2}}{s_1}\overline{B}_2(t){\color{red}.} \label{senso}
\end{equation}
 Moreover, we may use (\ref{Bi avg}) to denote the \textit{dimensionless} average concentration, as it is of the same form in both the dimensionless and dimensional contexts.

Applying the law of mass action to (\ref{rxns}) gives the kinetics equations:
\begin{subequations}
\begin{align}
	&\begin{aligned}
		\frac{\p B_1}{\p t} &= (1-B_\Sigma)C_1(x,0,t)- {}_1K_{\rd}B_1 - {}^1_2K_{\ra}B_1C_2(x,0,t) + {}^1_2K_{\rd}B_{12}, \label{react B1}              
	\end{aligned}\\
	&\begin{aligned}
		\frac{\p B_{12}}{\p t} &={}_2^1K_{\mathrm{a}}B_1C_2(x,0,t)-{}_2^1K_{\mathrm{d}}B_{12}+{}_1^2K_{\mathrm{a}}B_2C_1(x,0,t)-{}_1^2K_{\mathrm{d}}B_{12},  \label{react B12}                         
	\end{aligned}\\
	&\begin{aligned}
		\frac{\p B_2}{\p t} &= {}_2K_{\ra}(1-B_\Sigma)C_2(x,0,t)- {}_2K_{\rd}B_2+ {}_1^2K_{\rd}B_{12}  - {}^2_1K_{\ra}B_2C_1(x,0,t),  \label{react B2}
	\end{aligned}\\
	&\mathbf{B}(x,0)=\mathbf{0}, \label{react ic}
\end{align}\label{react Bsys}%
\end{subequations}
which hold on the reacting surface when $y=0$ and $x\in[0,1]$.  In (\ref{react ic}), $\mathbf{B}=(B_1,\ B_{12},\ B_2)^T$ is a vector in $\mathbb{R}^{3}$ whose components contain the three bound state concentrations.  In addition, the terms in equations (\ref{react B1})--(\ref{react B2}) have been ordered in accordance with Figures \ref{Figure: direct binding} and \ref{Figure: ligand switching}.

Edwards has shown \citep{edwards1999estimating} that transport effects dominate in a thin boundary layer near the reacting surface where diffusion and convection balance.  Hence the governing equations for $C_i$ are
\begin{subequations}
\begin{align}
& D_{\mathrm{r}} \frac{\p^2 C_1}{\p \eta^2}=\eta \frac{\p C_1}{\p x}, \label{react c1}\\
&  \frac{\p^2 C_2}{\p \eta^2}=\eta \frac{\p C_2}{\p x}.\label{react c2}
\end{align}
In (\ref{react c1})--(\ref{react c2}): $\eta=\Pe^{1/3}y$ is the boundary layer variable,  $\Pe=VH^2/(LD_2)\gg 1$ is the P{\'e}clet number, and $V$ is the characteristic velocity associated with our flow.  

Since $C_1$ is used up in the production of $B_1$ and $B_{12}$, and $C_2$ is used up in the production of $B_{12}$ and $B_2$, we have the diffusive flux conditions:
\begin{align}
&\frac{\p C_1}{\p \eta}(x,0,t)=\frac{\Da}{F_{\mathrm{r}}} \left(  \frac{\p B_1}{\p t}+\frac{\p B_{12}}{\p t}\right)\label{diff flux 1},\\
&\frac{\p C_2}{\p \eta}(x,0,t)=\Da \left(  \frac{\p B_{12}}{\p t}+\frac{\p B_{2}}{\p t}\right). \label{diff flux 2}
\end{align}
Equations (\ref{react c1})--(\ref{diff flux 2}) reflect the fact that in the boundary layer $C$ is in a quasi-steady-state where change is driven solely by the surface reactions (\ref{diff flux 1})--(\ref{diff flux 2}). Then, given the inflow and matching conditions
\begin{align}
C_i(0,\eta,t)=1, \label{layer inflow}\\
\lim_{\eta\to\infty}C_i(x,\eta,t)=1,\label{match}
\end{align}\label{layer C}
\end{subequations}
the solution to (\ref{layer C}) is given by
\begin{subequations}
\begin{align}
&C_1(x,0,t)=1-\frac{D_{\mathrm{r}}^{1/3}\Da}{F_{\mathrm{r}}\Gamma(2/3)3^{1/3}}\int_0^x\!\left(\frac{\p B_1}{\p t}+\frac{\p B_{12}}{\p t}\right)(x-\nu, t)\frac{\D \nu}{\nu^{2/3}}, \label{C1 int}\\
&C_2(x,0,t)=1-\frac{\Da}{\Gamma(2/3)3^{1/3}}\int_0^x\!\left(\frac{\p B_{12}}{\p t}+\frac{\p B_{2}}{\p t}\right)(x-\nu, t)\frac{\D \nu}{\nu^{2/3}}. \label{C2 int}
\end{align}\label{c ints}\end{subequations}
See \citep{edwards1999estimating} for details of a similar calculation.  During the injection phase, the bound state concentration is then governed by (\ref{react Bsys}) using (\ref{c ints}).  

In (\ref{diff flux 1})--(\ref{diff flux 2}) and (\ref{c ints})
\begin{equation}
\Da=\frac{{}_1k_{\mathrm{a}}R_{\mathrm{T}}(HL)^{1/3}}{(VD^{2})^{1/3}}
\end{equation} 
is the Damk{\"o}hler number--a key dimensionless parameter which measures the speed of reaction relative to the  transport into the surface.  In the experimentally relevant parameter regime of $\Da\ll 1$, the time scale for transport into the surface is much faster than the time scale for reaction.  In this case there is a only a weak coupling between the two processes, and (\ref{c ints}) shows that the unbound concentration at the surface is only a perturbation away from uniform inlet concentration.  When $\Da\to 0$ in (\ref{react Bsys}) using (\ref{c ints}), one recovers the well-stirred approximation in which transport into the surface  completely decouples from reaction.

On the other hand, when $\Da=O(1)$ the two processes occur on the same time scale, and \textit{ligand depletion effects} become more evident.  This is a phenomenon in which ligand molecules are transported into the surface to bind with receptor sites upstream, before they bind with receptor sites downstream.  Mathematically, this is reflected in the convolution integrals in (\ref{c ints}).  When $x\ll 1$ the convolution integral influences the unbound concentration at the surface less than when $x$ is larger.

\begin{figure}[tbhp!]
\centering
\begin{minipage}{.45\textwidth}
  \centering
  \includegraphics[width=6cm]{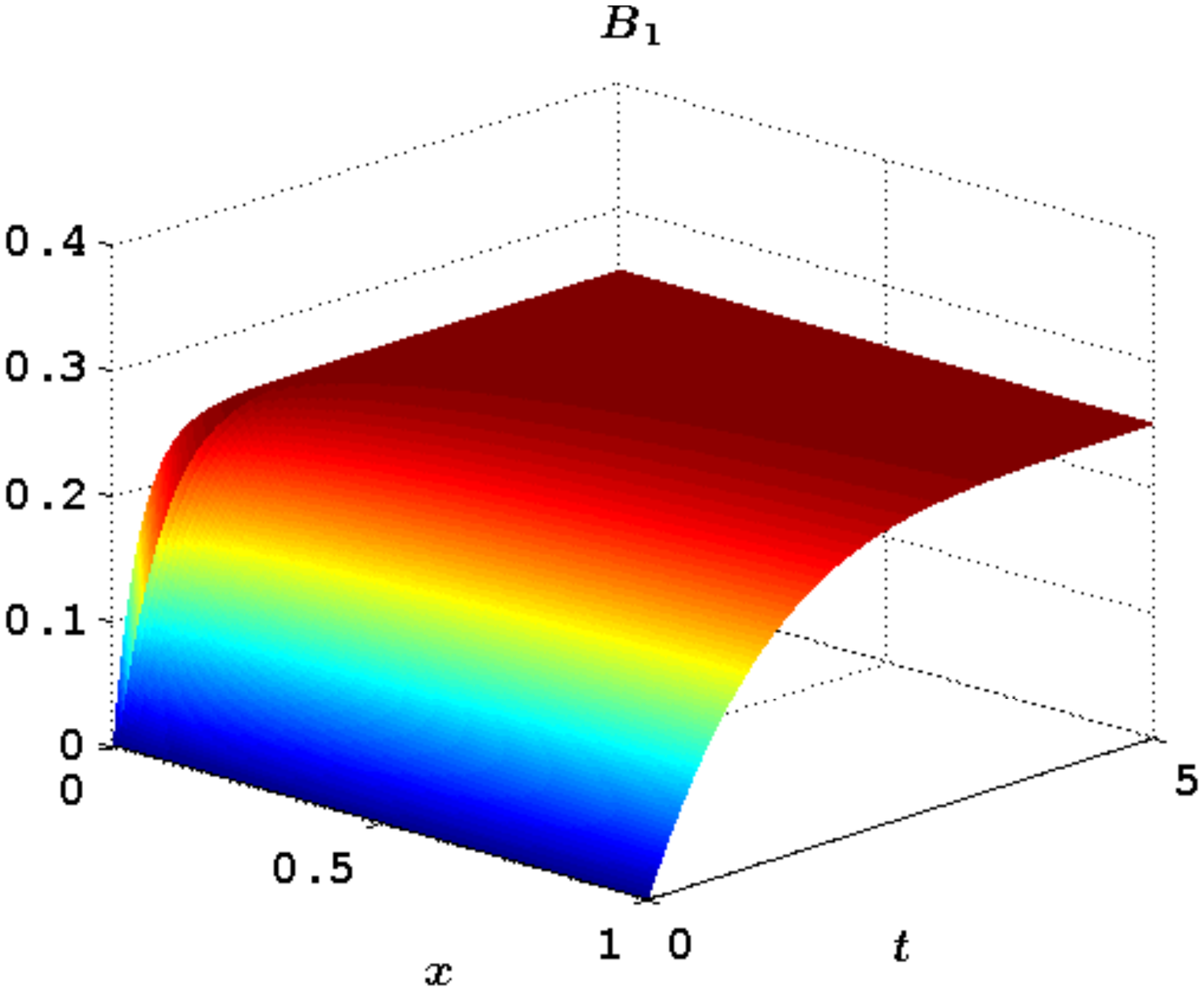}
\end{minipage}
\begin{minipage}{.45\textwidth}
  \centering
  \includegraphics[width=6cm]{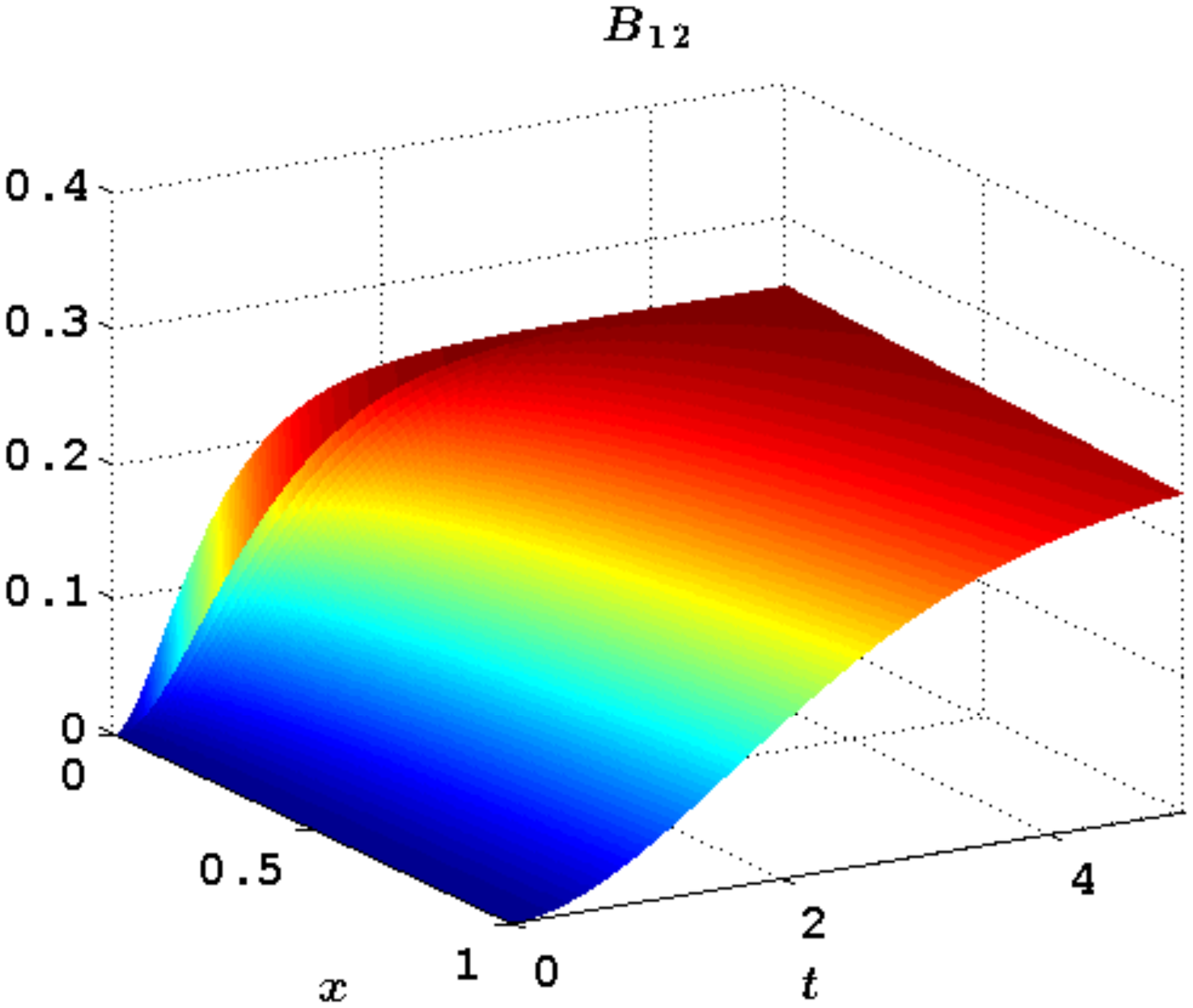}
\end{minipage}\vspace{-4cm}
\begin{minipage}{.5\textwidth}
  \centering
  \includegraphics[width=6cm]{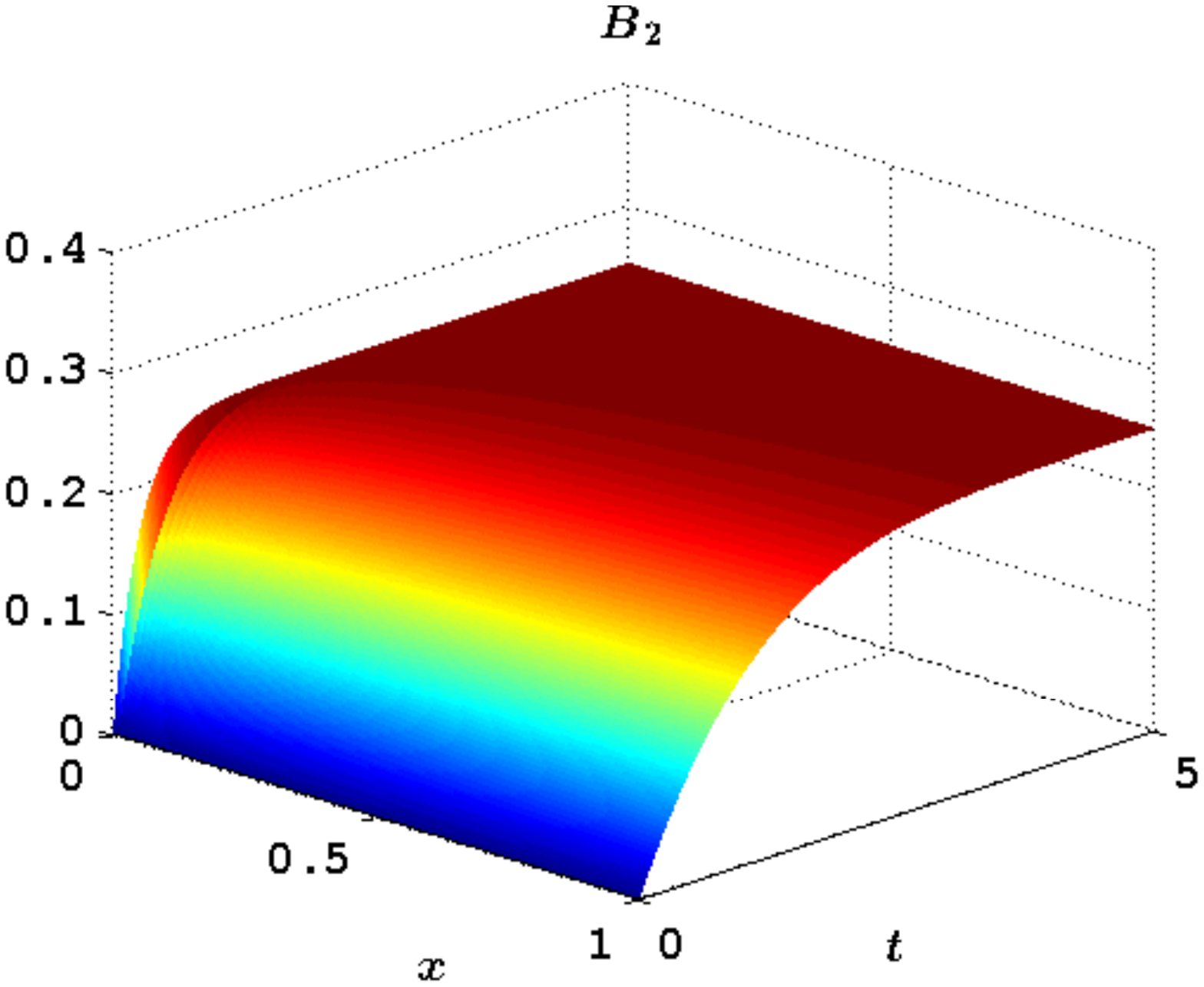}
\end{minipage}\vspace{-2cm}
\caption{Injection phase of biosensor experiment, up to $t=5$, obtained through solving (\ref{react Bsys}), (\ref{c ints}) with the numerical method described in Section \ref{Section: Numerics}. 
 All rate constants were taken equal to one, and $\Da$ taken equal to two to visualize upstream ligand depletion, which is especially evident in $B_{12}$.} \label{Figure: fd pics 1}
\end{figure}

A sample space-time curve for each of the reacting species concentrations $B_i(x,t)$ is depicted in Figure \ref{Figure: fd pics 1}, where we have shown the results of our numerical simulations described in Section \ref{Section: Numerics}. 
The $x$-axis represents the sensor, and $t$-axis represents time.  Injection begins at $t=0$, and ligand molecules bind with receptor sites as they are transported into the surface.  Binding proceeds as the injection continues; finally each of the concentrations achieve a chemical equilibrium in which there is a balance between association and dissociation.  Observe the spatial heterogeneity present in each of the bound state concentrations--the reaction proceeds faster near the inlet at $x=0$ than the rest of the surface.  This is precisely the ligand depletion phenomenon described in the above paragraph, and is particularly evident in the surface plot of $B_{12}$.  This is because in this simulation we have taken all of the rate constants equal to one, and either $EL_1$ or $EL_2$ must be present in order for $EL_1L_2$ to form.  Thus, in this case $EL_1L_2$ experiences effectively twice the ligand depletion of the other reacting species.

Furthermore, one may notice an apparent discontinuity in each of the surface plots depicted in Figure \ref{Figure: fd pics 1}--this reflects the weakly singular nature of the functions which we are attempting to approximate.  When $x\ll 1$, one may show $\mathbf{B}$ has the perturbation expansion
\begin{equation}
\mathbf{B}(x,t)={}^0\mathbf{B}(t)+\Da\ x^{1/3}\cdot {}^1\mathbf{B}(t)+ O(\Da^2\ x^{2/3})
\end{equation}%
(this is simply (\ref{b exp}) for $x\ll 1$).  It therefore follows that
\begin{equation}
\frac{\p \mathbf{B}}{\p x
}(x,t)=\frac{\Da\ {}^1\mathbf{B}(t)}{3x^{2/3}}+O\left( \frac{\Da^2}{x^{1/3}}\right).
\end{equation}
Hence, although the function $\mathbf{B}$ is well-defined and continuous near $x=0$, it has a vertical tangent at $x=0$.  The weakly-singular nature of $\mathbf{B}$ is magnified since $\Da=2$.  To resolve this region, one may think to 
adaptively change $\Delta x$ with the magnitude of $\p \mathbf{B}/\p x$.  However, because the sensogram reading $S(t)$ is computed over the region $[x_{\mathrm{min}},x_{\mathrm{max}}]$, we are not concerned with resolving this region and  a uniform step size is sufficient.  Moreover, our convergence results in Subsection \ref{Subsection: convergence study} demonstrate that a lack of resolution at $x=0$ does not affect our results in the region of interest $[x_{\mathrm{min}},x_{\mathrm{max}}]$.

\subsection{Wash Phase}
We now summarize the relevant equations for the wash phase.  In practice the injection phase is run until the bound state concentration reaches a steady-state
 \citep{richco2009sensor}.  This implies that because the bound ligand concentration evolves on a much slower time scale than the unbound ligand concentration \citep{edwards1999estimating}, the
 unbound ligand concentration will have also reached steady-state by the time the wash phase begins.  In particular, the unbound concentration on the surface will be uniform by the time the wash phase starts--\textit{i.e.}, $C_i(x,0,0)=1$.   Thus, the kinetics equations are given by (\ref{react Bsys}), with (\ref{react ic}) replaced by the steady solution to (\ref{react Bsys}) during the injection phase:
 \begin{subequations}
\begin{align}
&\mathbf{B}(x,0)=A^{-1}\mathbf{f},\label{ic: diss}\\
&A=\begin{pmatrix}
(1+\Kdo+\Kato) & 1 -\Kdto & 1\\
-\Kato & (\Kdto+\Kdot) & -\Kaot \\
\Kat & \Kat-\Kdot & (\Kat +\Kdt+\Kaot)
\end{pmatrix} ,\label{A mat}\\ &\fv=\begin{pmatrix}
1\\
0\\
\Kat
\end{pmatrix}.\label{f vec}
\end{align} \label{ic eqs: diss}%
\end{subequations}
Equations similar to (\ref{layer C}) hold:
\begin{subequations}
\begin{align}
C_i(0,\eta,t)=0, \label{layer inflow: diss}\\
\lim_{\eta\to\infty}C_i(x,\eta,t)=0.\label{match: diss}
\end{align}\label{c bc: diss}\noindent Equation (\ref{layer inflow: diss}) is the inflow condition, and (\ref{match: diss}) expresses the requirement that the concentration in the boundary layer must match the concentration $C_i(x,y,t)=0$ in the outer region.  Moreover, as in the injection phase one can use (\ref{react c1})--(\ref{diff flux 2}) together with (\ref{c bc: diss}) to show:
 \end{subequations}\begin{subequations}
\begin{align}
&C_1(x,0,t)=-\frac{D_{\mathrm{r}}^{1/3}\Da}{F_{\mathrm{r}}\Gamma(2/3)3^{1/3}}\int_0^x\!\left(\frac{\p B_1}{\p t}+\frac{\p B_{12}}{\p t}\right)(x-\nu, t)\frac{\D \nu}{x^{2/3}}, \label{C1 int: diss}\\
&C_2(x,0,t)=-\frac{\Da}{\Gamma(2/3)3^{1/3}}\int_0^x\!\left(\frac{\p B_{12}}{\p t}+\frac{\p B_{2}}{\p t}\right)(x-\nu, t)\frac{\D \nu}{\nu^{2/3}}. \label{C2 int: diss}
\end{align}\label{c ints: diss}\end{subequations}
Thus, during the wash phase the bound state evolution is governed by the (\ref{react B1})--(\ref{react B2}), (\ref{ic eqs: diss}), and (\ref{c ints: diss}).

\section{Effective Rate Constant Approximation} \label{Section: Effective Rate Constant Approximation}

During both phases of the experiment, the bound state concentration $\mathbf{B}(x,t)$ obeys a nonlinear set of IDEs which is hopeless to solve in closed form. However, we are 
ultimately interested in the average concentration $\overline{\mathbf{B}}(t)$, rather than the spatially-dependent function $\mathbf{B}(x,t)$, since from $\overline{\mathbf{B}}(t)$ we can construct the sensogram signal (\ref{senso}) (the quantity of interest).  Thus, we seek to find an approximation to $\overline{\mathbf{B}}(t)$, and begin by
 finding one during the injection phase.  We first average each side of (\ref{react Bsys}), with $C_1(x,0,t)$ and $C_2(x,0,t)$ given by (\ref{c ints}), in the sense of (\ref{Bi avg}).  Immediately, we are
 confronted with terms such as
\begin{align}
&\overline{B_1C_2}=\overline{B_1\left(1-\frac{\Da}{3^{1/3}\Gamma(2/3)}\int_0^x\!\left(\frac{\p B_{12}}{\p t}+\frac{\p B_2}{\p t} \right)\frac{\D \nu}{(x-\nu)^{2/3}}\right)}, \label{nonlinear avg}
\end{align}
on the right hand side of (\ref{react B1}).   In the experimentally relevant case of small $\Da$, we are motivated to expand ${\bf B}(x,t)$ in a perturbation series:
\begin{equation}
{\bf B}(x,t)={}^0{\bf B}(x,t)+O(\Da).
\label{bser}
\end{equation}
In this limit, the leading order of (\ref{c ints}) is just $C_i=1$.  Using this result in (\ref{react Bsys}), we have that the governing equation for ${}^0{\bf B}$ is independent of $x$:
$$
{\mathrm{d}\,{}^0{\bf B}\over \mathrm{d}t}=-A\,{}^0{\bf B}+\mathbf{f},
$$
where $A$ is given by (\ref{A mat}) and $\mathbf{f}$ by (\ref{f vec}).  Hence the leading-order approximation
\begin{equation}
\pre{0}{}{\mathbf{B}}(t)=A^{-1}(I-\rme^{-At})\mathbf{f}
\label{b0sol}
\end{equation}
is independent of space. Substituting \eqref{b0sol} into (\ref{c ints}), the time-dependent terms may be factored out of the integrand, 
leaving the spatial dependence of $C_j$ varying as $x^{1/3}$.  This is the only spatial variation in (\ref{react Bsys}) at $O(\Da)$; hence we may write
\begin{equation}
\mathbf{B}(x,t)={}^0\mathbf{B}(t)+\Da\ x^{1/3}\cdot{}^1\mathbf{B}(t)+O(\Da^2). \label{b exp}
\end{equation}
As a result of (\ref{b exp}) we have the relation
\begin{equation}
\Da\ B_i(x,t)=\Da \pre{0\!}{}{B}_i(t)+O(\Da^2), \label{trick}
\end{equation}
which may be used to show  the right hand side of (\ref{nonlinear avg}) is equal to
\begin{align}
&\overline{B}_1-\Da\ \overline{h}\cdot \pre{0\!}{}{B}_{1}\left(\frac{\D \pre{0\!}{}{B}_{12}}{\D t}+\frac{\D \pre{0\!}{}{B}_2}{\D t}\right)+O(\Da^2),\label{nonlinear avg 2}\\
&h(x)=\frac{3^{2/3}x^{1/3}}{\Gamma(2/3)}.\nonumber
\end{align}
We then average (\ref{trick}), and use the resulting relation in (\ref{nonlinear avg 2}) to show the right hand side of (\ref{nonlinear avg}) reduces to:
\begin{align*}
&\overline{B_1 C_2}=\overline{B}_1\left[ 1-\Da\ \overline{h}\left(\frac{\D \overline{B}_{12}}{\D t}+\frac{\D \overline{B}_2}{\D t}\right)\right]+O(\Da^2).
\end{align*}
In this manner, we can derive a set of nonlinear ODEs for $\overline{\mathbf{B}}(t)$ of the form:
\begin{subequations}
\begin{align}
&\frac{\D \overline{\mathbf{B}}}{\D t}=M^{-1}(\overline{\mathbf{B}})(-A\overline{\mathbf{B}}+\fv)+O(\Da^2), \label{ml erc eq}\\
&\overline{\mathbf{B}}(0)=\mathbf{0}, \label{ml erc ic}
\end{align}
where
\begin{align}
&M(\overline{\mathbf{B}})=I+\Da\ N(\overline{\mathbf{B}}),\label{mass matrix}\\
&N(\overline{\mathbf{B}})=\begin{pmatrix}
\frac{D_{\mathrm{r}}^{1/3}\overline{h}}{F_{\mathrm{r}}} (1-\avBs)  & \frac{D_{\mathrm{r}}^{1/3}\overline{h}}{F_{\mathrm{r}}} (1-\avBs)  - \Kato\overline{h} \cdot \avBo&-\Kato \overline{h}\cdot\avBo\\
\Kato\overline{h}\cdot\avBo&\Kato\overline{h}\cdot\avBo+\Kaot \left(\frac{D_{\mathrm{r}}^{1/3}\overline{h}}{F_{\mathrm{r}}}  \right)\avBt  & \Kaot\left( \frac{D_{\mathrm{r}}^{1/3}\overline{h}}{F_{\mathrm{r}}}\right) \avBt \\
-\Kato \left( \frac{D_{\mathrm{r}}^{1/3}\overline{h}}{F_{\mathrm{r}}} \right)\avBt & -\Kato\left(\frac{D_{\mathrm{r}}^{1/3}\overline{h}}{F_{\mathrm{r}}}  \right)\avBt+\Kat\overline{h}(1-\avBs) & \Kat\overline{h}(1-\avBs)
\end{pmatrix}.
\end{align}\label{ml erc group}\end{subequations}
We have also derived a set of ERC equations for the wash phase, they take take the form:
\begin{subequations}
\begin{align}
&\frac{\D \overline{\mathbf{B}}}{\D t}=M^{-1}(\overline{\mathbf{B}})(-\mathcal{D}\overline{\mathbf{B}})+O(\Da^2), \label{ml erc eq: diss}\\
&\overline{\mathbf{B}}(0)=A^{-1}\mathbf{f}, \label{ml erc ic: diss}\\
&\mathcal{D}=\begin{pmatrix}
\Kdo & -\Kdto & 0\\
0 & \Kdot+\Kdto  & 0\\
0 & -\Kdot & \Kdt
\end{pmatrix}\label{diss op},
\end{align}\label{ml erc group: diss}
where $M(\overline{\mathbf{B}})$ is as in (\ref{mass matrix})  
\end{subequations}

Following \citep{edwards2002testing}, we refer to the Ordinary Differential Equation (ODE) systems  (\ref{ml erc group}) and (\ref{ml erc group: diss}) as our \textit{Effective Rate Constant} (ERC) \textit{Equations}.  A significant advantage of our ERC equations is that these ODEs are far easier to solve numerically than their IDE counterparts.  To solve (\ref{ml erc group}) or (\ref{ml erc group: diss}), one may simply apply their linear multistage or multistep formula of choice.  This feature renders our ERC equations attractive for data analysis, since they can be readily implemented into a regression algorithm when attempting to determine the rate constants associated with the reactions (\ref{rxns}).  Since experimental data is still forthcoming, we do not employ a regression algorithm to fit the rate constants in (\ref{ml erc group}) and (\ref{ml erc group: diss}) to biosensor data. Synthetic data for the kinetic rate constants was used in our numerical simulations.

\begin{figure}[tbhp!]
\centering
\begin{minipage}{.48\textwidth}
  \centering
  \includegraphics[width=6cm]{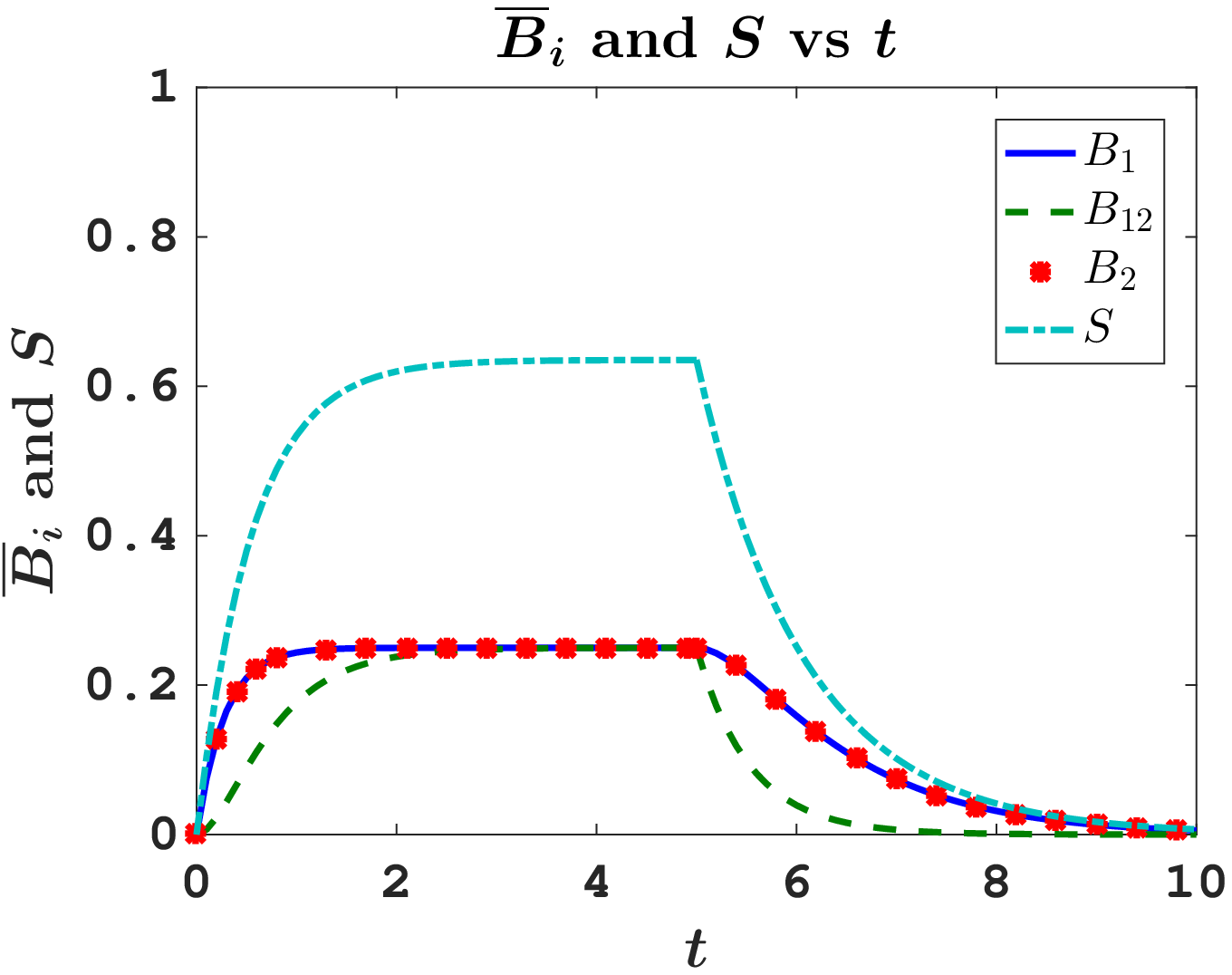}
\end{minipage}
\begin{minipage}{.48\textwidth}
  \centering
  \includegraphics[width=6cm]{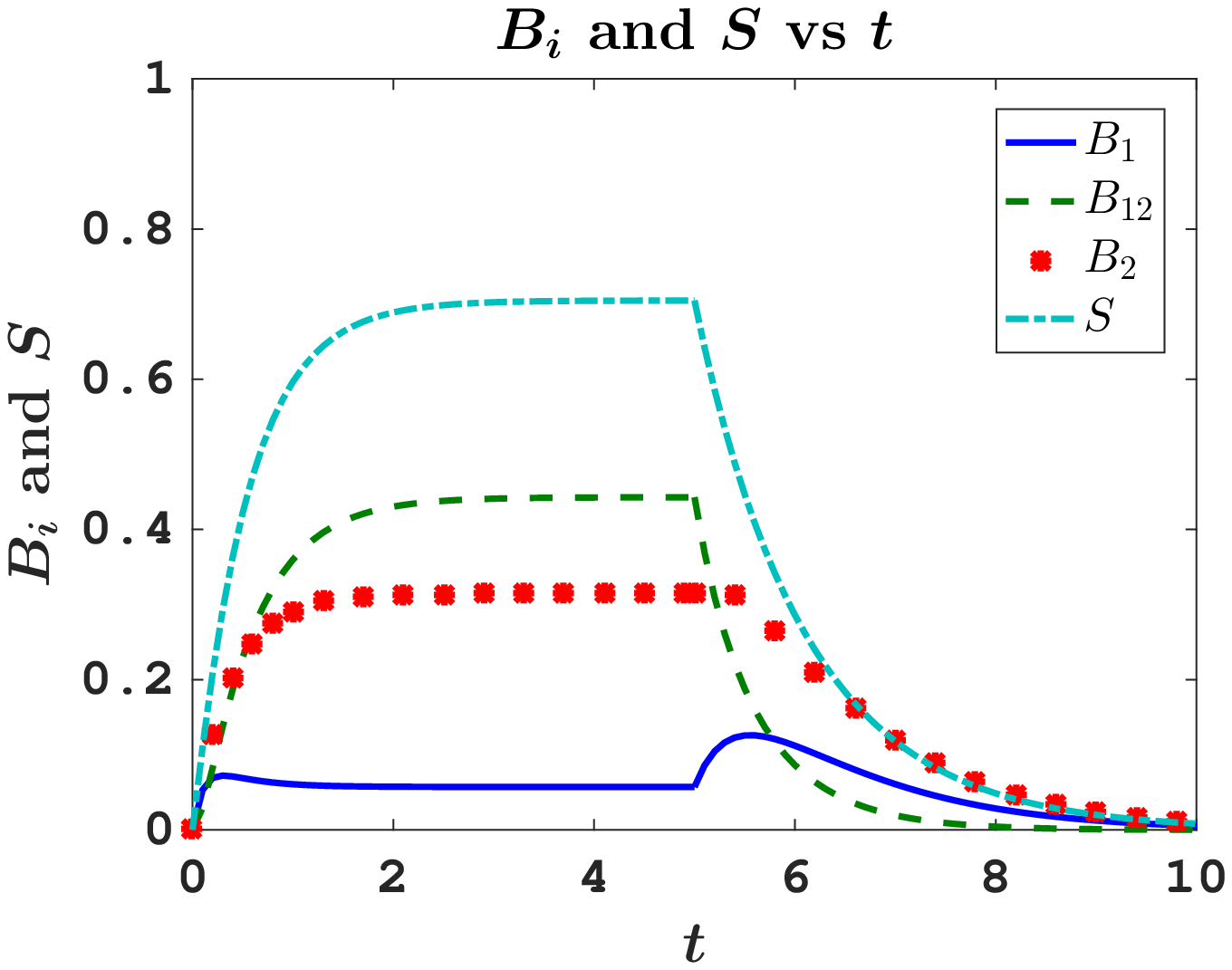}
\end{minipage}
\caption{Left:  The solution of injection phase ERC equations (\ref{ml erc group}) is depicted from $t=0$ to $t=5$, and the solution of the wash phase ERC equations (\ref{ml erc group: diss}) is depicted from $t=5$ to $t=10$. Here all of the rate constants were taken equal to 1.  Right:  The injection phase (\ref{ml erc group}) is depicted from $t=0$ to $t=5$, and the wash phase (\ref{ml erc group: diss}) is depicted from $t=5$ to $t=10$. Here all of the rate constants were taken equal to 1 except ${}_2^1K_{\mathrm{a}}=10$.   Both: The Damk{\"o}hler number was taken equal to $\Da=0.1$.}
\label{Figure: ERC sol}
\end{figure}

Solutions of our ERC equations for different parameter values are depicted in Figure \ref{Figure: ERC sol}.
First consider the solutions depicted on the left.  Here the injection phase (\ref{ml erc group}) has been run from $t=0$ to $t=5$ and the wash phase (\ref{ml erc group: diss}) has been run from $t=5$ to $t=10$.  Furthermore, all rate constants were taken equal to one and the Damk{\"o}hler number was $\Da=0.1$.  During the injection phase it is seen that $\overline{B}_1$ and $\overline{B}_2$ reach equilibrium after approximately one second, while $\overline{B}_{12}$ takes approximately two seconds.  This is not a surprise: we are injecting equal amounts of both ligands, all the rate constants are the same, and either $EL_1$ or $EL_2$ must already be present in order for $EL_1L_2$ to form.  The equality of the rate constants is also the reason why all three species attain the same steady-state.  Mathematically, the steady-state of $\mathbf{B}$ during the injection phase is given by (\ref{ic: diss}), and one can readily verify that $A^{-1}\mathbf{f}=(1/4,\ 1/4,\ 1/4	)^T$ when all of the rate constants are equal to one.  Physically, each of the species  ultimately achieves the same balance between association and dissociation.  Furthermore, the fact that all of the rate constants are the same is the reason why  $\overline{B}_{12}$ decays to zero faster than the other two species: $EL_1L_2$ transitions to either $EL_1$ or $EL_2$ at the same rate as the latter two species transition into an empty receptor $E$.

Now consider the solutions depicted on the right in Figure \ref{Figure: ERC sol}.  As with the previous case the injection phase has been run from $t=0$ to $t=5$ and the wash phase has been run from $t=5$ to $t=10$.  However this time, all the rate constants have been taken equal to one except ${}_2^1K_{\mathrm{a}}$, which was taken equal to ${}_2^1K_{\mathrm{a}}=10$. During the injection phase it is seen that $\overline{B}_1$ quickly reaches a local maximum, and then decreases to steady-state. Since ${}_2^1K_{\mathrm{a}}$ is an order of magnitude larger than the other rate constants, after a short period of time $L_2$ molecules bind with $EL_1$ at a faster rate than $L_1$ molecules bind with empty receptor sites. This results in the chemical equilibrium between $EL_1$ and $EL_1L_2$ depicted on the right in Figure \ref{Figure: ERC sol}.  From these observations it is clear why the steady-state value of $\overline{B}_{12}$ is larger than the previous case.  However, it may be counterintuitive to observe that  $\overline{B}_2$ reaches a larger steady-state value in the solutions depicted on the right than the solutions depicted to the left.  Although one may think the vast majority $L_2$ molecules should be used in forming $EL_1L_2$, the increase in $EL_1L_2$ also increases the concentration of empty receptor sites. The continuous injection of $L_2$ therefore drives the average concentration $\overline{B}_2$ to a larger steady-state value.  During the wash phase, it is seen that $\overline{B}_1$ reaches a global maximum after approximately $t\approx5.75$ seconds.  The increase in $\overline{B}_1$ during the wash phase is a direct consequence of $L_2$ molecules dissociating from $EL_1L_2$.  Since only pure buffer is flowing through the biosensor during the wash phase, it is seen in Figure \ref{Figure: ERC sol} that each of the average concentrations $\overline{B}_i$ decay to zero.





\section{Numerics} \label{Section: Numerics}

To verify the $O(\Da^2)$ accuracy of our ERC approximation derived in Section \ref{Section: Effective Rate Constant Approximation}, we now develop a numerical approximation to the IDE system (\ref{react Bsys}), where $C_1(x,0,t)$ and $C_2(x,0,t)$ are given by (\ref{c ints}).  We focus on the injection phase, since the wash phase is similar. Our approach is based on the numerical method described in \citep{edwards2002testing}.  Semi-implicit methods have been previously used with great
success to solve reaction-diffusion equations \citep{nie2006efficient}, as they are typically robust, efficient, and accurate.  Similarly, in our problem we exploit the structure of the integrodifferential operator, which naturally suggests a semi-implicit method in time.  Moreover, since our method is semi-implicit in time we avoid the expense and complication of solving a nonlinear system at each time step. Convergence properties and remarks concerning stability, are discussed in Subsection \ref{Subsection: convergence study};  however,  we first turn our attention to deriving our numerical method in Subsection \ref{Subsection: semi-implicit finite difference algorithm}.

\subsection{Semi-implicit finite difference algorithm}\label{Subsection: semi-implicit finite difference algorithm}

We discretize the spatial interval $[0,1]$ by choosing $N+1$ equally spaced discretization nodes $x_i=i\Delta x$, for $i=0,\ \ldots, N$, and discretize time by setting $t_n=n\Delta t $, for $n=0,\ \ldots$. Having chosen our discretization nodes and time steps, we seek to discretize (\ref{react Bsys}), where $C_1(x,0,t)$ and $C_2(x,0,t)$ are given by (\ref{c ints}).  Note that this requires discretizing both the time derivatives and the convolution integrals; we first turn our attention to the latter, and focus on $C_1(x,0,t)$.  We would like to apply the trapezoidal rule  to spatially discretize (\ref{C1 int}), however the integrand of $C_1(x,0,t)$ is singular when $\nu=0$.  To handle the singularity we subtract and add
\begin{equation}
\left(\frac{\p B_1}{\p t}+\frac{\p B_{12}}{\p t}\right)(x-\nu,t)|_{\nu=0}\label{trick 2}
\end{equation}
from the integrand.  Doing so yields
\begin{align}
\begin{aligned}
C_1(x,0,t)=&1-\frac{D_{\mathrm{r}}^{1/3}\Da}{F_{\mathrm{r}} 3^{1/3}\Gamma(2/3)}\ \bigg\{\int_0^x\bigg[\bigg(\frac{\p B_1}{\p t}+\frac{\p B_{12}}{\p t}\bigg)(x-\nu,t)\\ 
&\ - \bigg(\frac{\p B_1}{\p t}+\frac{\p B_{12}}{\p t}\bigg)(x,t)\bigg]\frac{\mathrm{d} \nu}{\nu^{2/3}}+3 x^{1/3}\bigg(\frac{\p B_1}{\p t}+\frac{\p B_{12}}{\p t}\bigg)(x,t)\bigg\},
\end{aligned}\label{reg C1}
\end{align}
where we have used the fact that (\ref{trick 2}) is independent of $\nu$.  Then choosing a discretization node $x=x_i$ and a time step $t=t_n$, we apply the trapezoidal rule to (\ref{reg C1}) to obtain
\begin{align}
\begin{aligned}
C_1(x_i,0,t_n)=&1-\frac{D_{\mathrm{r}}^{1/3}\Da}{F_{\mathrm{r}} 3^{1/3}\Gamma(2/3)}\ \bigg\{0\cdot \frac{\dx}{2}+\sum_{j=1}^{i-1}\bigg[\bigg(\frac{\p B_1}{\p t}+\frac{\p B_{12}}{\p t}\bigg)(x_i-x_j,t_n)\\ 
&\ - \bigg(\frac{\p B_1}{\p t}+\frac{\p B_{12}}{\p t}\bigg)(x_i,t_n)\bigg]\frac{\Delta x}{x_j^{2/3}}+\bigg[\bigg(\frac{\p B_1}{\p t}+\frac{\p B_{12}}{\p t}\bigg)(0,t_n)\\
&\ -\bigg(\frac{\p B_1}{\p t}+\frac{\p B_{12}}{\p t}\bigg)(x_i,t_n) \bigg]\frac{\dx}{2 x_i^{2/3}}+3 x^{1/3}_i\bigg(\frac{\p B_1}{\p t}+\frac{\p B_{12}}{\p t}\bigg)(x_i,t_n)\bigg\},
\end{aligned}\label{disc C1 2}
\end{align}
when $x_i>0$; simply evaluating (\ref{reg C1}) at $x=x_0$ gives  $C(x_0,0,t_n)=1$.  The first term in the sum is zero, because in a similar manner to Appendix  B of \citep{zumbrum2013extensions} we have
\begin{equation}
\lim_{\nu\to 0}\bigg(\frac{\p B_k}{\p t}(x-\nu, t_n) -\frac{\p B_k}{\p t}(x,t)\bigg)\frac{1}{\nu^{2/3}}=\lim_{\nu\to 0}\nu^{1/3}\bigg(\frac{\p B_k}{\p t}(x-\nu, t_n) -\frac{\p B_k}{\p t}(x,t)\bigg)\frac{1}{\nu},
\end{equation}
which implies
\begin{equation}
\lim_{\nu\to 0}\bigg(\frac{\p B_k}{\p t}(x-\nu, t_n) -\frac{\p B_k}{\p t}(x,t)\bigg)\frac{1}{\nu^{2/3}}=\lim_{\nu\to 0}\nu^{1/3}\frac{\p^2 B_k}{\p x\p t}(x,t)=0,\label{B reg}
\end{equation}
for $k=1,\ 12, $ or $2$.  The last equality follows since we expect $\p B_i/\p t$ to be sufficiently regular for fixed $x>0$.  The expansion (\ref{b exp}) shows that this is certainly true when $\Da\ll 1$, however when $\Da=O(1)$ or larger the nonlinearity in (\ref{react Bsys}) renders any analytic approximation to $B_i$ beyond reach.  Our results in Subsection \ref{Subsection: convergence study} show that our method indeed converges when $\Da=O(1)$ or larger.

We now turn our attention to discretizing the time derivatives.  We denote our approximation to $B_j(x_i,t_n)$ by 
\begin{equation}
B_j(x_i,t_n)\approx B^j_{i,n},
\end{equation}
and approximate the time derivatives through the formula
\begin{equation}
\frac{\p B_j}{\p t}(x_i,t_n)\approx \frac{B^j_{i,n}-B^j_{i,n-1}}{\Delta t}:=\frac{\Delta B^j_{i,n}}{\Delta t}.\label{time derivative approx}
\end{equation}
Our approximation (\ref{time derivative approx}) holds for all reacting species $j=1,\ 12,\ 2$, each of our discretization nodes $x_i$, and each time step $t_n$.  As we shall show below, we treat $\Delta B^j_{i,n}$ as separate variable used to update $B^j_{i,n}$ at each iteration of our algorithm.

With our time derivatives discretized as (\ref{time derivative approx}), the fully-discretized version of $C_1(x,0,t)$ is given by substituting (\ref{time derivative approx}) into (\ref{disc C1 2}):\begin{subequations}
\begin{align}
\begin{aligned}
C^1_{i,n}=&1-\frac{D_{\mathrm{r}}^{1/3}\Da}{F_{\mathrm{r}} 3^{1/3}\Gamma(2/3)}\ \bigg\{\sum_{j=1}^{i-1}\bigg[\bigg(\frac{\Delta B^1_{i-j,n}}{\Delta t}+\frac{\Delta B^{12}_{i-j,n}}{\Delta t}\bigg)- \bigg(\frac{\Delta B^1_{i,n}}{\Delta t}+\frac{\Delta B^{12}_{i,n}}{\Delta t}\bigg)\bigg]\frac{\Delta x}{x_j^{2/3}}\\ & \ +\bigg[\bigg(\frac{\Delta B^1_{0,n}}{\Delta t}+\frac{\Delta B^{12}_{0,n}}{\Delta t}\bigg) -\bigg(\frac{\Delta B^1_{i,n}}{\Delta t}+\frac{\Delta B^{12}_{i,n}}{\Delta t}\bigg)\bigg]\frac{\dx}{2 x_i^{2/3}}+3 x^{1/3}_i\bigg(\frac{\Delta B^1_{i,n}}{\Delta t}+\frac{\Delta B^{12}_{i,n}}{\Delta t}\bigg)\bigg\},
\end{aligned}\label{disc C1 3}
\end{align}
for $i>0$, and $C^1_{0,n}=1$.  The function $C_{2}(x,0,t)$ has a similar discretization which we denote as $C_2(x_i,0,t_n)\approx C^2_{i,n}$.  Thus, our numerical method takes the form:
\begin{align}
&\frac{\Delta B^{1}_{i,n+1}}{\dt}=(1-B^{\Sigma}_{i,n})C^1_{i,n+1}-\Kdo B^{1}_{i,n}-\Kato B^{1}_{i,n}C^2_{i,n+1}+\Kdto B^{12}_{i,n},\\
& \frac{\Delta B^{12}_{i,n+1}}{\dt}=\Kato B^1_{i,n}C^2_{i,n+1}-\Kdto B^{12}_{i,n}+\Kaot B^2_{i,n}C^{1}_{i,n+1}-\Kdot B^{12}_{i,n},\\
&\frac{\Delta B^{2}_{i,n+1}}{\dt}=\Kat(1-B^{\Sigma}_{i,n})C_{i,n+1}^2-\Kdt B^{2}_{i,n}+\Kdot B^{12}_{i,n}-\Kaot B^{2}_{i,n}C^1_{i,n+1}.
\end{align}
\label{num method}%
\end{subequations}
We enforce the initial condition (\ref{react ic}) at our $N+1$ discretization nodes through the condition $B^j_{i,0}=0$ for $j=1,\ 12,\ 2$, and $i=1,\ldots, N$.  Observe that our method (\ref{num method}) is semi-implicit rather than fully-implicit.  This renders (\ref{num method}) linear in $\Delta B^{j}_{i,n+1}$, and as a result we can write
\begin{subequations}
\begin{equation}
\frac{\Delta \mathbf{B}_{i,n+1}}{\dt}=M^{-1}_{i,n}(\mathbf{B}_{i,n})(A^{-1}_{i,n+1}\mathbf{B}_{i,n}+\mathbf{f}_{i,n+1}),  \label{num method 2}
\end{equation}
where $\mathbf{B}_{i,n}=(B^{1}_{i,n},\ B^{12}_{i,n},\ B^2_{i,n})^T$.  Hence, by using a method which is only semi-implicit in time we  avoid the expense and complication of solving a nonlinear system at each time step.  Having solved for $\Delta \mathbf{B}_{i,n+1}$ using (\ref{num method 2}), we march forward in time at a given node $x_i$ through the formula
\begin{equation}
\mathbf{B}_{i,n+1}=\mathbf{B}_{i,n}+\frac{1}{2}(3\Delta \mathbf{B}_{i,n+1}-\Delta \mathbf{B}_{i,n}),\label{AB2}
\end{equation}
\label{num method recipe}%
\end{subequations}
which is analogous to a second-order Adams-Bashforth formula.  

In addition, we chose a method that is implicit in $C_1(x,0,t)$ and $C_2(x,0,t)$ also due to the form of the convolution integrals.  From (\ref{c ints}) we see  $C_1(x,0,t)$ and $C_2(x,0,t)$ depend on $B_j(\nu,t)$ only for $\nu\le x$.  Thus by choosing a method that is implicit in $C_1(x,0,t)$ and $C_2(x,0,t)$, we are able to use the \textit{updated values} of $B_j(x,t)$ in the convolution integrals by first computing the solution at $x=0$, and marching our way downstream at each time step.


To make this notion more precise we note that in (\ref{num method 2}) the matrix $M^{-1}_{i,n}(\mathbf{B}_{i,n})$ depends only upon $\mathbf{B}_{i,n}$, however because of the convolution integrals $C^1_{i,n+1}$ and $C^2_{i,n+1}$, the matrix $A_{i,n+1}$ and vector $\mathbf{f}_{i,n+1}$ depend upon $\mathbf{B}_{l,n+1}$ for $l<i$.  Thus, at each time step $n+1$ we first determine $\mathbf{B}_{0,n+1}$.  Next, we increment $i$ and use the value of $\mathbf{B}_{0,n+1}$ in (\ref{num method recipe}) to determine $\mathbf{B}_{1,N+1}$.  We proceed by iteratively marching our way downstream from $x_2$ to $x_N$ to determine $\mathbf{B}_{2,n+1},\ \ldots,\ \mathbf{B}_{N,n+1}$.  Intuitively, the updated information from the convolution integral flows downstream from left to right at each time step.  We may repeat this procedure for as many time steps as we wish.  In addition, we remark that the formula (\ref{num method}) was initialized with one step of Euler's method.  

Furthermore, with our finite difference approximation to $\mathbf{B}(x,t)$, we can determine the average quantity 
\begin{equation}
\overline{B}(t)=(\overline{B}_1(t),\ \overline{B}_{12}(t),\ \overline{B}_2(t))^T
\end{equation}
with the trapezoidal rule
\begin{equation}
\overline{\mathbf{B}}(t_n)\approx\frac{1}{x_{\mathrm{max}}-x_{\mathrm{min}}}\left(\frac{\dx}{2} \mathbf{B}_{m,n}+\dx\sum_{i=m+1}^{M-1}\mathbf{B}_{i,n}+\frac{\dx}{2}\mathbf{B}_{M,n}\right).\label{avg approx}
\end{equation}
In (\ref{avg approx}), the indices $i=m$ and $i=M$ correspond to $x_{\mathrm{min}}=m\dx$ and $x_{\mathrm{max}}=M\dx$.  Our nodes were chosen to align with $x_{\mathrm{min}}$ and $x_{\mathrm{max}}$ to avoid interpolation error.

\subsection{Convergence study} \label{Subsection: convergence study}

\subsubsection{Spatial Convergence}\label{Subsubsection: Spatial convergence}

We now examine the spatial rate of convergence of our numerical method.  Since from $\overline{\mathbf{B}}$ we can compute the quantity of interest (\ref{senso}), we derive estimates for the rate at which our numerical approximation converges to $\overline{\mathbf{B}}$.  Furthermore, because the system (\ref{react Bsys}), (\ref{c ints}) is nonlinear, our analysis will focus on the experimentally relevant case of $\Da\ll 1$.  In addition, we will derive estimates only for the injection phase of the experiment, since the wash phase is similar.

To proceed, we consider the average variant of (\ref{react Bsys}), (\ref{c ints}).  Averaging (\ref{react Bsys}) in the sense of (\ref{Bi avg}) gives:
\begin{subequations}
\begin{align}
	&\begin{aligned}
		\frac{\mathrm{d} \overline{B}_1}{\mathrm{d} t} &= \overline{(1-B_\Sigma)C_1(x,0,t)}- {}_1K_{\rd}\overline{B}_1  - {}^1_2K_{\ra}\overline{B_1C_2(x,0,t)}+ {}^1_2K_{\rd}\overline{B}_{12} ,
	  \label{react B1 error}              
	\end{aligned}\\
	&\begin{aligned}
		\frac{\mathrm{d} \overline{B}_{12}}{\mathrm{d} t} &= {}^1_2K_{\ra}\overline{B_1C_2(x,0,t)}- {}^1_2K_{\rd}\overline{B}_{12} + {}^2_1K_{\ra}\overline{B_2C_1(x,0,t)} 
  - {}_1^2K_{\rd}\overline{B}_{12},  \label{react B12 error}                         
	\end{aligned}\\
	&\begin{aligned}
		\frac{\mathrm{d} \overline{B}_2}{\mathrm{d} t} &= {}_2K_{\ra}\overline{(1-B_\Sigma)C_2(x,0,t)}- {}_2K_{\rd}\overline{B}_2+ {}_1^2K_{\rd}\overline{B}_{12}  - {}^2_1K_{\ra}\overline{B_2C_1(x,0,t)}.  \label{react B2 error}
	\end{aligned}\\
	&\overline{\mathbf{B}}(0)=\mathbf{0}. \label{react ic}
\end{align}\label{react Bsys error}%
\end{subequations}
As in Subsection \ref{Subsection: semi-implicit finite difference algorithm}, we handle the singularity in (\ref{C1 int}) by adding and subtracting (\ref{trick 2}) from the integrand of (\ref{C1 int}) to write $C_1(x,0,t)$ as in (\ref{reg C1}).  The unbound ligand concentration $C_2(x,0,t)$ has a representation analogous to (\ref{reg C1}).  In the following analysis we limit our attention to (\ref{react B1 error}), since the analysis for equations (\ref{react B12 error})--(\ref{react B2 error}) is nearly identical.

We proceed by anaylzing each of the terms in (\ref{react B1 error}):
\begin{subequations}
\begin{align}
&-\Kdo\overline{B}_1,\label{lin term 2}\\
&\Kdto\overline{B}_{12},\label{lin term 1}\\
&\overline{C_{1}(x,0,t)},\label{C1 err}\\
&-\overline{B_{\Sigma}C_1(x,0,t)},\label{nonlin C1 err}\\
&-\Kato \overline{B_1 C_2(x,0,t)}\label{nonline C2 err}.
\end{align}
\end{subequations}
Upon inspecting (\ref{reg C1}) and using linearity of the averaging operator, we see that three terms contribute to (\ref{C1 err}):\begin{subequations}
\begin{align}
&\overline{1}\label{const err}\\
&-\overline{\frac{D_{\mathrm{r}}^{1/3}\Da}{F_{\mathrm{r}}3^{1/3}\Gamma(2/3)}\int_0^x\bigg(\frac{\p B_1}{\p t}+\frac{\p B_{12}}{\p t}\bigg)(x-\nu,t)-\bigg(\frac{\p B_1}{\p t}+\frac{\p B_{12}}{\p t}\bigg)(x,t)\frac{\mathrm{d}\nu}{\nu^{2/3}}},\label{avg int 1}\\
&\overline{3x^{1/3}\bigg(\frac{\p B_1}{\p t}+\frac{\p B_{12}}{\p t}\bigg)}.\label{x lin}
\end{align}\end{subequations}
In a similar manner, (\ref{nonlin C1 err}) and (\ref{nonline C2 err}) each imply that we incur error from the terms:\begin{subequations}
\begin{align}
&- \overline{B}_{\Sigma}\label{avg Bsig}\\ 
&\overline{\frac{D_{\mathrm{r}}^{1/3}\Da B_{\Sigma}}{F_{\mathrm{r}}3^{1/3}\Gamma(2/3)}\int_0^x\bigg(\frac{\p B_1}{\p t}+\frac{\p B_{12}}{\p t}\bigg)(x-\nu,t)-\bigg(\frac{\p B_1}{\p t}+\frac{\p B_{12}}{\p t}\bigg)(x,t)\frac{\mathrm{d}\nu}{\nu^{2/3}}},\label{avg int 2}\\
&-\overline{3x^{1/3}B_{\Sigma}\bigg(\frac{\p B_1}{\p t}+\frac{\p B_{12}}{\p t}\bigg)},\label{nonlinear term 3}\\
&-\overline{\Kato B}_{1}\label{avg B1}\\
&\overline{\frac{\Kato \Da B_{1}}{3^{1/3}\Gamma(2/3)}\int_0^x\bigg(\frac{\p B_{12}}{\p t}+\frac{\p B_{2}}{\p t}\bigg)(x-\nu,t)-\bigg(\frac{\p B_{12}}{\p t}+\frac{\p B_{2}}{\p t}\bigg)(x,t)\frac{\mathrm{d}\nu}{\nu^{2/3}}},\label{avg int 3}\\
&-\overline{3\Kato x^{1/3}B_{1}\bigg(\frac{\p B_{12}}{\p t}+\frac{\p B_{2}}{\p t}\bigg)}.\label{nonlinear term 4}
\end{align}\end{subequations}

Let us denote the trapezoidal rule of a function $f(x)$ over the interval $[a,b]$ by $\mathcal{T}(f(x),[a,b])$.  Then since $\mathcal{T}(1,[x_{\mathrm{min}},x_{\mathrm{max}}])$ is exact, the term (\ref{const err}) does not contribute to the spatial discritization error.

Next we decompose the expansion (\ref{b exp}) into its individual components to obtain
\begin{equation}
B_j(x,t)={}^0B_j(t)+\Da\ x^{1/3}\cdot {}^1B_j(t)+O(x^{2/3}\Da^2),\label{b exp comp}
\end{equation}  
for $j=1,\ 12,\ 2$.  Substituting (\ref{b exp comp}) into (\ref{lin term 1}), (\ref{lin term 2}), (\ref{avg Bsig}) (\ref{avg B1}),  and using the fact that $\mathcal{T}(x^{1/3},[x_{\mathrm{min}},x_{\mathrm{max}}])$ converges at a rate $O(\dx^2)$ shows that each of these terms converge at a rate of $O(\Da\dx^2)$.  Similarly, one can substitute (\ref{b exp comp}) into (\ref{x lin}), (\ref{nonlinear term 3}), and (\ref{nonlinear term 4}), and use the fact that $\mathcal{T}(x^{1/3},[x_{\mathrm{min}},x_{\mathrm{max}}])$ converges at a rate of $O(\dx^2)$, to show that each of these terms converge at a rate of of $O(\Da \dx^{2})$.  

It remains to determine the error associated with (\ref{avg int 1}), (\ref{avg int 2}), and (\ref{avg int 3}), so we turn our attention to (\ref{avg int 1}) and substitute (\ref{b exp comp}) into (\ref{avg int 1}) to obtain
\begin{equation}
-\frac{D_{\mathrm{r}}^{1/3}\Da}{F_{\mathrm{r}}3^{1/3}\Gamma(2/3)(x_{\mathrm{max}}-x_{\mathrm{min}})}\bigg(\frac{\mathrm{d} {}^1B_1}{\mathrm{d} t}+\frac{\mathrm{d}{}^1B_{12}}{\mathrm{d}t}\bigg)(t)\int_{x_{\mathrm{min}}}^{x_{\mathrm{max}}}\int_0^x(x-\nu)^{1/3}-x^{1/2})\frac{\mathrm{d}\nu}{\nu^{2/3}}, \label{avg int 1: v2}
\end{equation}
where we have used the definition of our averaging operator (\ref{Bi avg}).  In writing (\ref{avg int 1: v2}), neglected higher-order terms which do not contribute to the leading-order spatial discretization error.  Since the coefficient of the integral in (\ref{avg int 1: v2}) is a function of time alone, this coefficient does not contribute to the leading-order spatial discretization error and we neglect it in our analysis.  Hence, to compute the spatial discretization error associated with (\ref{avg int 1: v2}), we calculate the error  associated with applying the trapezoidal rule to the double integral
\begin{equation}
\int_{x_{\mathrm{min}}}^{x_{\mathrm{max}}}\int_0^x[(x-\nu)^{1/3}-x^{1/3}]\frac{\mathrm{d}\nu}{\nu^{2/3}}\ \mathrm{d}x. \label{avg int 1: v3}
\end{equation}
Treating the inner integral as a function of $x$ we define
\begin{equation}
f(x)=\int_0^x \!\left( (x-\nu)^{1/3}-x^{1/3}\right)\nu^{-2/3}\ \D \nu,\label{f int}
\end{equation}
whose closed form is given by
\begin{equation}
f( x)=\frac{ x^{2/3}}{2} \left(\frac{2^{1/3} \sqrt{\pi } \Gamma \left(\frac{1}{3}\right)}{\Gamma \left(\frac{5}{6}\right)}-6\right)\label{exact f}.
\end{equation}
Towards applying the trapezoidal rule to (\ref{avg int 1: v3}), we first note $\mathcal{T}(f,[0,x_i])$ converges at a rate of $O(\dx^{4/3})$.  This is seen by first rewriting (\ref{f int}) as
\begin{equation}
 \int_0^{\Delta x}\! [(x-\nu)^{1/3}-x^{1/3}]\nu^{-2/3}\ \D \nu+\int_{\Delta x}^{ x_i}\! [(x-\nu)^{1/3}-x^{1/3}]\nu^{-2/3}\ \D \nu.
\end{equation}
The term on the right converges at a rate of $O(\dx^2)$, and the term on the left converges at a rate of $O(\dx^{4/3})$, which follows from expanding $(x-\nu)^{1/3}$ about $\nu=0$, and using the definition of the trapezoidal rule.

Applying the trapezoidal rule to (\ref{avg int 1: v3}) then gives 
\begin{align}
\begin{aligned}
&\Da^2\int_{\xmin}^{\xmax}\int_0^{x}\!\left((x-\nu)^{1/3}-x^{1/3}\right) \nu^{-2/3}\ \D \nu\ \D x\\ &\quad =\frac{\Da^2\dx}{2}\mathcal{T}(f(x),[0,x_m])+\sum_{i=m+1}^{M-1}\Da^2\dx \mathcal{T}(f(x),[0,x_i]) \\ &\qquad+\frac{\Da^2\dx}{2}\mathcal{T}(f(x),[0,x_M])+O(\Da^2\dx^{2}),
\end{aligned}
\end{align}
where we have let $x_{\min}=x_m=m\dx$, and $x_{\max}=x_M=M\dx$.  Since $\mathcal{T}(f,[0,x_i])$ converges at a rate of $O(\dx^{4/3})$, the right hand side of the above is 
\begin{align}
\begin{aligned}
&\left(\frac{\Da^2\dx}{2}f(x_m)+O(\Da^2\dx^{7/3})\right)+\sum_{i=m+1}^{M-1}(\Da^2\dx f(x_i)+O(\Da^2\dx^{7/3}))\\&\quad+\left(\frac{\Da^2\dx}{2}f(x_M)+O(\Da^2\dx^{7/3})\right)+O(\Da^2\dx^{2}).
\end{aligned}
\end{align}
To compute our results in Section \ref{Section: Effective Rate Constant Approximation Verification}, we took $\xmin=0.2,\ \xmax=0.8$, in accordance with the literature \citep{edwards2002testing}.  Hence, in the above sum there are approximately $0.6N=0.6\dx^{-1}$ terms on the order of $O(\Da^2\dx^{7/3})$, and the above sum reduces to
\begin{align} \begin{aligned}
&\Da^2 \left(\frac{\dx}{2}f(x_m)+\sum_{i=m+1}^{M-1}\dx f(x_i)
 +\frac{\dx}{2}f(x_M)\right)+O(\Da^2\dx^{4/3})\\&\quad+O(\Da^2\dx^{5/3}).\end{aligned}\label{final err}
\end{align}
The dominant error in (\ref{final err}) is $O(\Da^2\dx^{4/3})$, thus the spatial discretization error associated with (\ref{avg int 1}) is $O(\Da^2\dx^{4/3})$.  When measuring convergence we used values of $\xmin=.25,\ \xmax=.75$ to facilitate progressive
 grid refinement; however, it is clear that these values of $\xmin$ and $\xmax$ do not change our argument.  A similar argument shows the spatial discretization error associated with the nonlinear terms (\ref{avg int 2}) and (\ref{nonlinear term 4}) is $O(\Da^2\dx^{4/3})$.   

\begin{figure}[tbhp!]
\vspace{-1cm}
\centering
  \includegraphics[width=9cm]{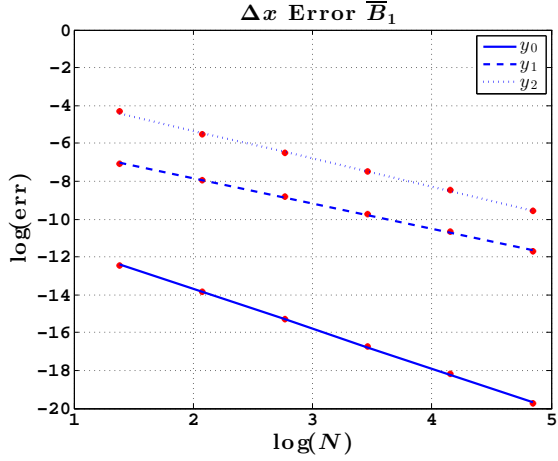}\vspace{-2cm}
\caption{Spatial convergence for $\overline{B}_1$, $\Da=.01,1,10$. The (solid) line $y_0=-2.10x-9.49$ was fit to the error when $\Da=.01$.  The (dashed) line $y_1=-1.33x-5.17$ was fit to the error when $\Da=1$.  The (dotted) line $y_2=-1.48x-2.36$ was fit to the error when $\Da=10$.  All three lines have an $R^2$ coefficient of $R^2=.99$.  The rate constants were taken equal to: $\Kdo=1,\ \Kat=1,\  \Kdt=1,\ \Kaot=1/2,\ \Kdot=2,\ \Kato=2,$ and $\Kdto=1/2$.}
\label{Figure: convergence fig}
\end{figure}

We have depicted our spatial convergence measurements for $\overline{B}_1$ in Figure \ref{Figure: convergence fig} and tabulated them in Table \ref{Table: convergence tab}.  To obtain these results, we first computed a reference solution, with $\dx=\dt=1/512$.  We then created a series of test solutions with mesh width $\dx=1/2^j$, for $j=2,\ldots, 7$, keeping $\dt=1/512$ constant. Next, we computed $\overline{\mathbf{B}}$ by averaging our reference solution and test solutions at each time step with the trapezoidal rule as in (\ref{avg approx}).  We then computed the error between each test solution and the reference solution by taking the maximum difference of the two over all time steps.

\begin{table}[tbhp!]

\centering
\caption{Convergence results for the reacting species.  Here $\Da=.01,\ 1, 10$.{\smallskip}}
{\makegapedcells \begin{tabular}{|c|c|c|c|}
\hline
        & $\Da\ll 1 $           &     $\Da=O(1)$  & $\Da\gg 1$\\ \hline
$\overline{B}_{1}$ & $O(\dx^{2.09})$  & $O(\dx^{1.33})$ & $O(\dx^{1.48})$\\ \hline
$\overline{B}_{12}$ & $O(\dx^{2.04})$  & $O(\dx^{1.42})$ & $O(\dx^{1.53})$\\ \hline
$\overline{B}_{2}$ & $O(\dx^{2.09})$  & $O(\dx^{1.33})$ & $O(\dx^{1.46})$\\ \hline
\end{tabular}}\label{Table: convergence tab}

\end{table}

From our results, we see that our method converges at a rate of $O(\dx^2)$ when $\Da\ll 1$, $O(\dx^{4/3})$ when $\Da=O(1)$, and $O(\dx^{3/2})$ when $\Da\gg 1$.  The reduction in convergence when $\Da$ increases from small to moderate may be attributed to the $O(\Da^2\dx^{4/3})$ contributions from (\ref{avg int 1}), (\ref{avg int 2}), and (\ref{avg int 3}).  There are
 two competing magnitudes of error in (\ref{react B1 error}): one of $O(\Da\ \dx^2)$ (from terms  (\ref{lin term 1}), (\ref{lin term 2}), (\ref{x lin}), (\ref{avg Bsig}), (\ref{nonlinear term 3}), (\ref{avg B1}), and (\ref{nonlinear term 4})), and one of $O(\Da^2 \dx^{4/3})$ (from the integral terms (\ref{avg int 1}), (\ref{avg int 2}), and (\ref{avg int 3})).  When $\Da^2\dx^{4/3}< \Da\ \dx^{2}$, or $\Da<\dx^{2/3}$, the former is larger.  Conversely, when $\dx^{2/3}<\Da$ the latter is larger.

When $\Da\gg 1$, the bound state evolves on a longer time scale of the form \citep{edwards1999estimating}
\begin{equation}
t_{\mathrm{w}}=t/\Da. \label{tw}
\end{equation}
In this case, the characteristic time scale for reaction for reaction is much faster than the characteristic time scale for transport into the surface, and one typically refers to  $\Da\gg 1$ as the \textit{transport-limited} regime.  Substituting (\ref{tw}) into (\ref{react Bsys}), (\ref{c ints}), one may find the leading-order approximation to the resulting system for $\Da\gg 1$ by neglecting the left hand side of (\ref{react B1})--(\ref{react B2}).  Doing so one finds that even a leading-order approximation to (\ref{react B1})--(\ref{react B2}) is nonlinear, rendering any error estimates in the transport-limited regime beyond reach. Nonetheless, our results in Figure \ref{Figure: convergence fig} and Table \ref{Table: convergence tab} show that convergence is not an issue when $\Da\gg 1$.

\subsubsection{Temporal Convergence}\label{Subsubsection: Temporal Convergence}

\begin{figure}[tbhp!]
\centering
  \includegraphics[width=8cm]{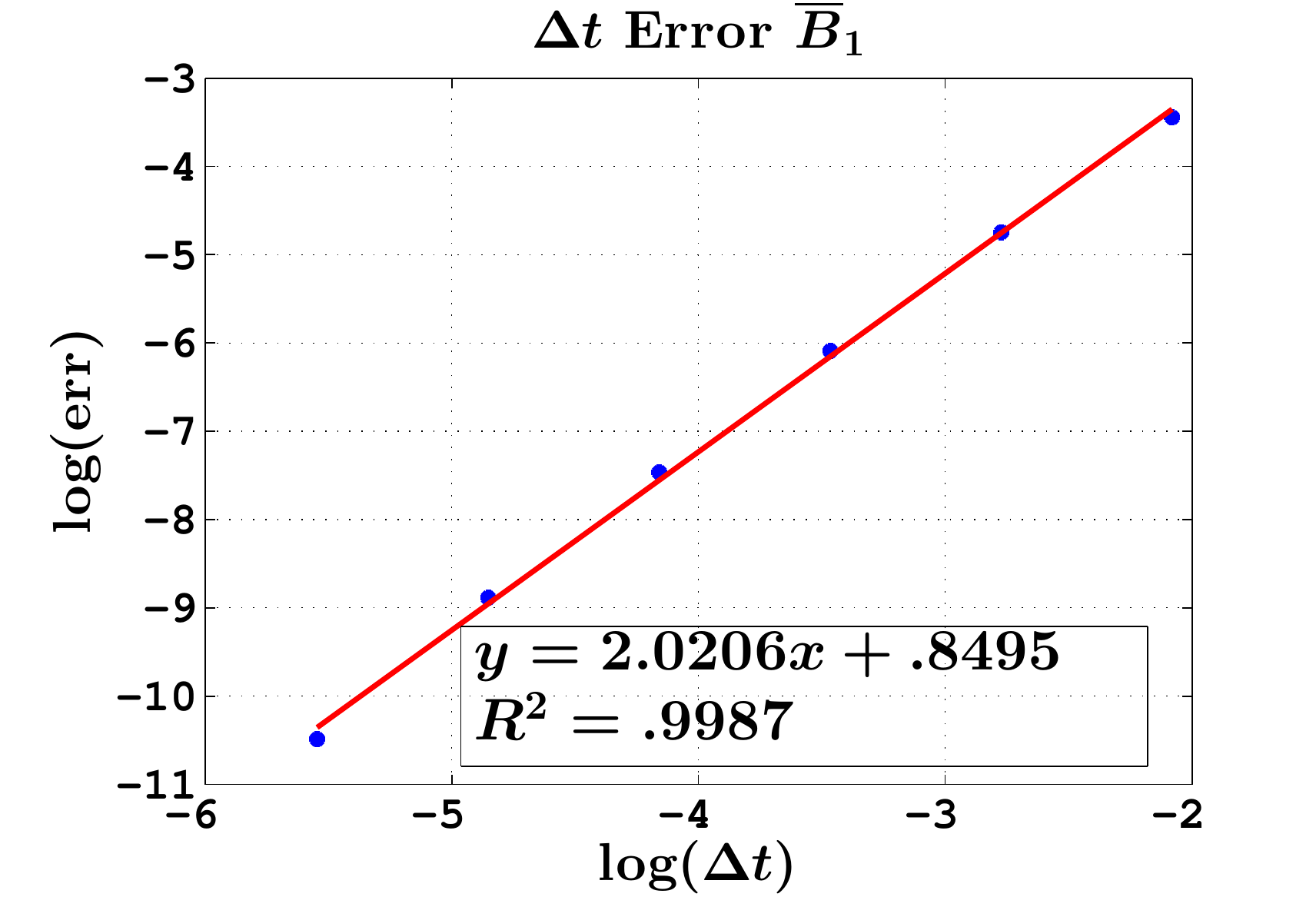}
\caption{Temporal convergence when $\Da=.01$, and $\Kdo=1,\ \Kat=1,\  \Kdt=1,\ \Kaot=1/2,\ \Kdot=2,\ \Kato=2,$ and $\Kdto=1/2$. As expected, our method converges at a rate on the order of $O(\dt^2)$.}
\label{Figure: temporal convergence}
\end{figure}

Since our time stepping method (\ref{AB2}) is analogous to a second-order Adams-Bashforth formula, we expect our method to achieve second-order accuracy in time.  Figure \ref{Figure: temporal convergence} shows that this is indeed the case when $\Da=.01$, and the rate constants are $\Kdo=1,\ \Kat=1,\  \Kdt=1,\ \Kaot=1/2,\ \Kdot=2,\ \Kato=2,$ and $\Kdto=1/2$ (as in Subsection \ref{Subsubsection: Spatial convergence} when measuring spatial convergence).  Temporal convergence was measured in an analogous manner to spatial convergence.

 However, we note that measuring temporal convergence when $\Da=O(1)$ is computationally prohibitive, since spatial convergence is $O(\Da^2\dx^{4/3})$ in this case, so in order for the spatial and temporal errors to balance one must have $O(\Da^2\dx^{4/3})=O(\dt^2)$ or $\dx=\dt^{3/2}$.  
 Nonetheless, our results from Section \ref{Section: Effective Rate Constant Approximation Verification} demonstrate that our finite difference approximation agrees with our ERC approximation for a wide parameter range, so we are not concerned with temporal convergence of our method when $\Da=O(1)$ or larger.

\subsubsection{Stability Remarks}\label{Subsubsection: Stability Remarks}

We now make brief remarks concerning the stability of our method.  Recall from Subsection \ref{Subsection: semi-implicit finite difference algorithm} we first determine the value of $B_i(x,t)$ upstream at $x=0$, and iteratively march our way downstream to $x=1$ at each time step.  Therefore, we expect any instabilities at $x=0$ to propagate downstream.  Requiring that there no instabilities at $x=0$ is equivalent to asking that our time stepping method (\ref{AB2}) is stable for the ODE system found by replacing $C_1(x,0,t)$ and $C_2(x,0,t)$ with the constant function 1 in (\ref{react Bsys}).    Though we do not have precise stability estimates for this system, numerical experimentation has shown that our time steps need to be sufficiently small in order to ensure that our numerical approximation  is well behaved.

\section{Effective Rate Constant Approximation Verification}\label{Section: Effective Rate Constant Approximation Verification}
With our numerical method in hand, we are now in a position to verify the accuracy of our ERC approximations (\ref{ml erc group}) and (\ref{ml erc group: diss}).    We tested the accuracy of our ERC equations when $\Da=0.1$, and $\Da=0.45$;  the results are below in Figure \ref{Figure: ERC accuracy da1} and Tables \ref{Table: ERC error}--\ref{Table: ERC error wash}.

\begin{figure}[tbhp!]
\centering
\begin{minipage}{.48\textwidth}
  \centering
  \includegraphics[width=6.25cm]{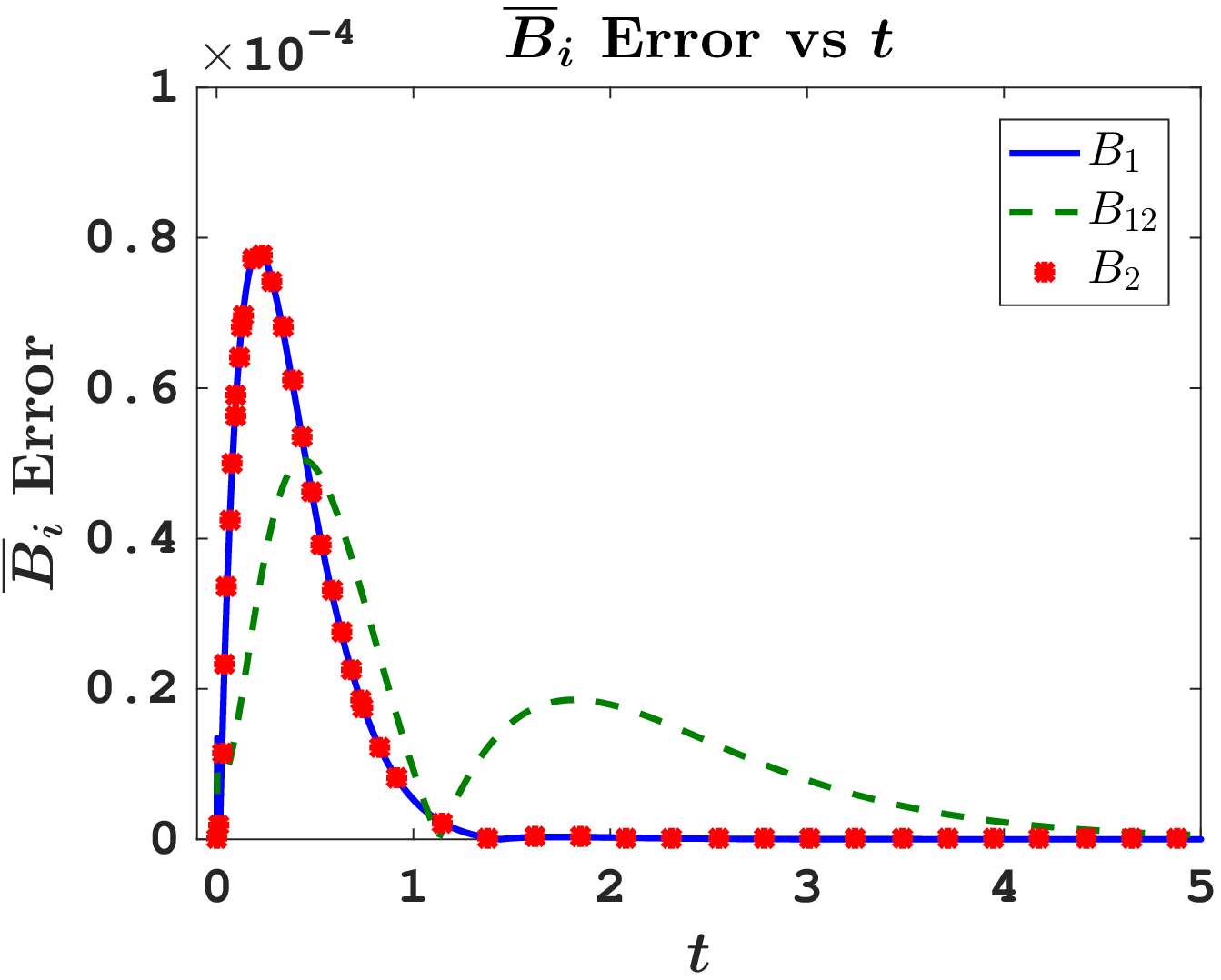}
\end{minipage}
\begin{minipage}{.48\textwidth}
  \centering
  \includegraphics[width=6.25cm]{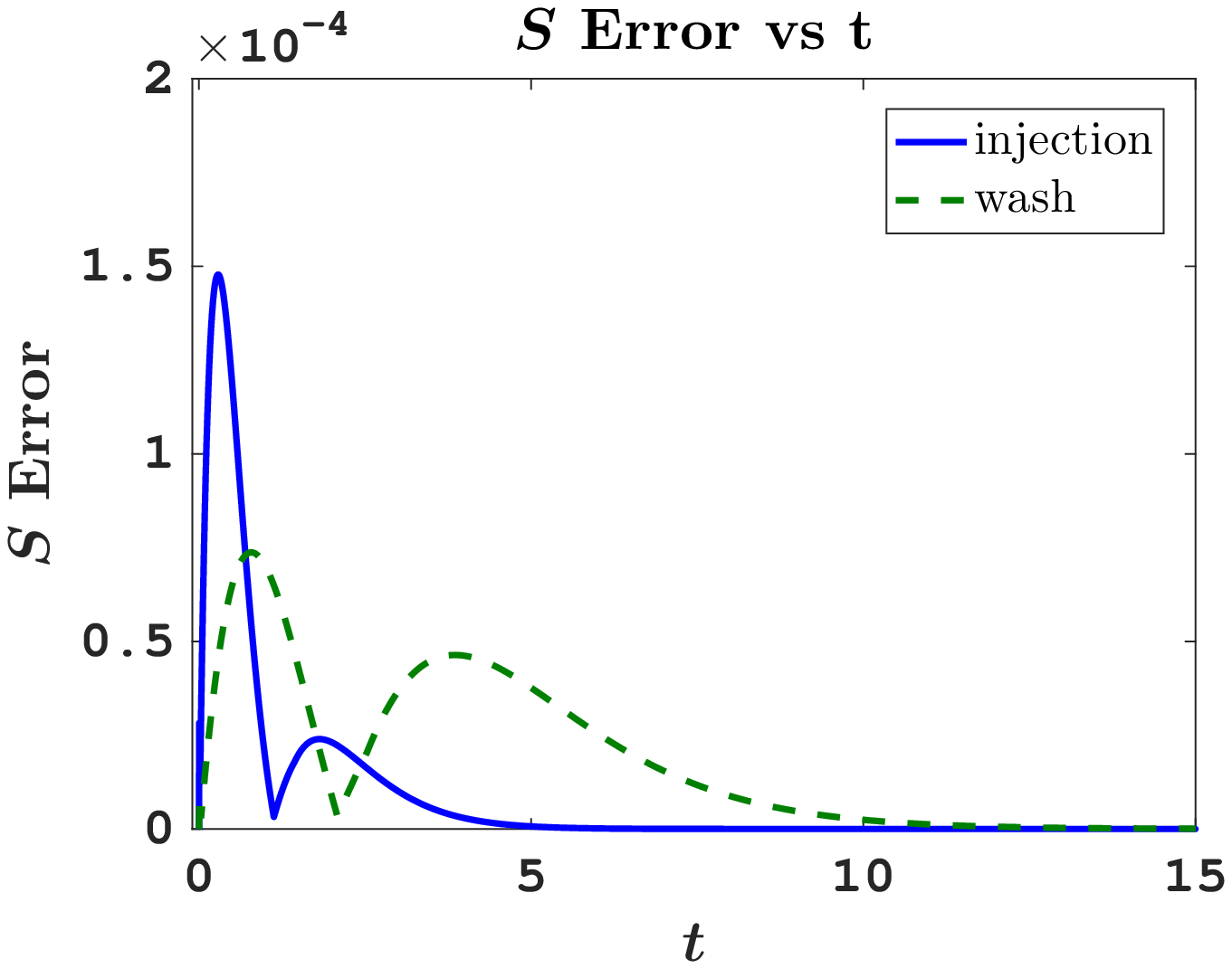}
\end{minipage}
\caption{Left:  Error in the reacting species concentrations during the injection phase, computed by taking the absolute difference between the solution of our ERC
 equation (\ref{ml erc group}) and our finite difference solution.  We have taken $\Da=0.1$, and all of the rate constants equal to 1.  Since the errors for $\overline{B}_1$ and $\overline{B}_2$ are identical we plotted the error for $\overline{B}_2$ on a coarser mesh, however the errors for all three species were computed on precisely the same time steps.  Right: the error in the sensogram signal during both phases, when $\Da=.1$ and all of the rate constants are taken equal to 1.}
\label{Figure: ERC accuracy da1}
\end{figure}

\begin{table}[tbhp!]

\centering
\caption{Maximum difference between our injection phase ERC approximation (\ref{ml erc group}) and our finite difference solution.  All rate constants take equal to 1 in both cases.{\smallskip}}
{\makegapedcells
\begin{tabular}{|l|c|c|c|c|}
\hline
        & $\overline{B}_1 $           &     $\overline{B}_{12}$  & $\overline{B}_2$ & $\mathcal{S}$\\ \hline
$\Da=.1$ & 7.81$\times10^{-5}$  & $5.04\times 10^{-5}$ & $7.81\times 10^{-5}$ & $1.47\times 10^{-4}$\\ \hline 
$\Da=.45$ & $1.00\times 10^{-3}$  & $6.93\times 10^{-4}$ &  $1.00\times 10^{-3}$ & $2.00\times10^{-3}$\\
\hline
\end{tabular}}\label{Table: ERC error}

\end{table}

\begin{table}[tbhp!]

\centering
\caption{Maximum difference between our wash phase ERC approximation (\ref{ml erc group: diss}) and our finite difference solution.  All rate constants take equal to 1 in both cases.{\smallskip}}
{\makegapedcells
\begin{tabular}{|l|c|c|c|c|}
\hline
        & $\overline{B}_1 $           &     $\overline{B}_{12}$  & $\overline{B}_2$ & $\mathcal{S}$\\ \hline
$\Da=.1$ & 3.578$\times10^{-5}$  & $3.40\times 10^{-5}$ & $3.58\times 10^{-5}$ & $7.37\times 10^{-5}$\\ \hline 
$\Da=.45$ & $4.33\times 10^{-4}$  & $4.62\times 10^{-4}$ &  $4.33\times 10^{-4}$ & $9.45\times10^{-4}$\\
\hline
\end{tabular}}\label{Table: ERC error wash}

\end{table}

\begin{figure}[tbhp!]
\centering
\begin{minipage}{.48\textwidth}
\centering
  \includegraphics[width=6cm]{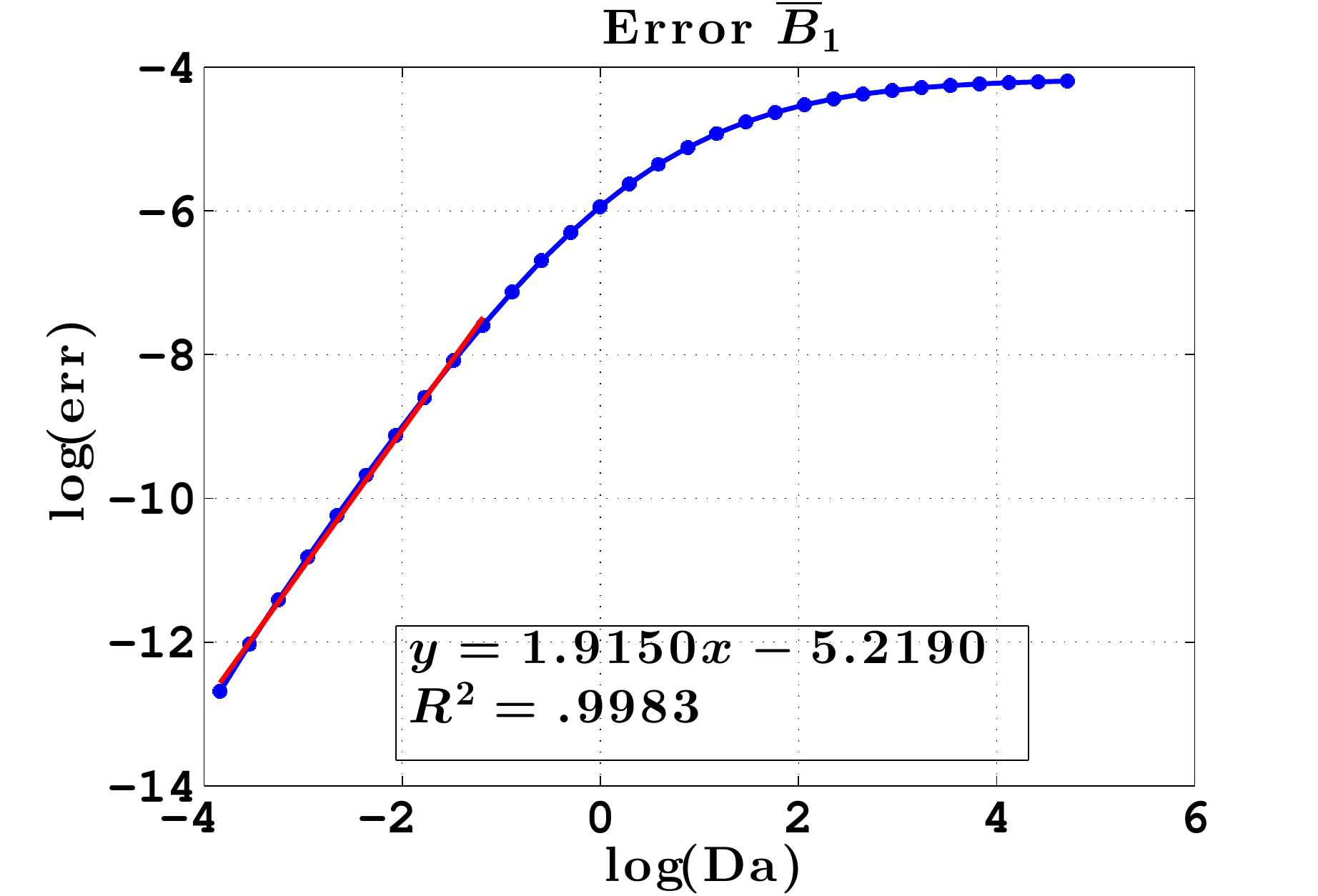}
\end{minipage}
\centering
\begin{minipage}{.48\textwidth}
\centering
  \includegraphics[width=6cm]{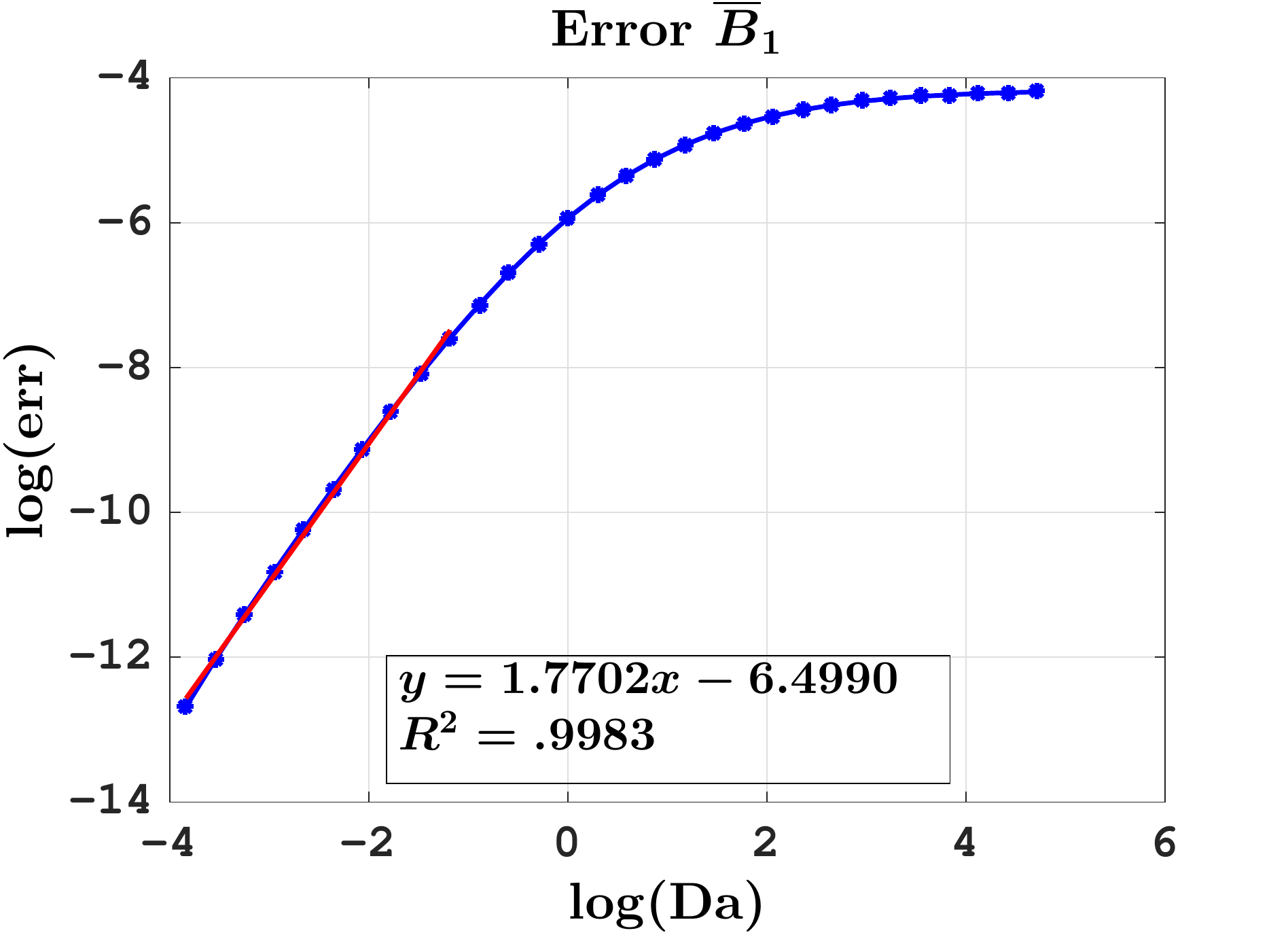}
 \end{minipage}
\caption{Left: Absolute error in our injection phase approximation (\ref{ml erc group}) over all time, for different values of $\Da$.  Right: absolute error in our wash phase approximation (\ref{ml erc group: diss}) over all time, for different values of $\Da$. Both:   the rate constants were taken equal to: $\Kdo=1/2,\Kat=1,\Kdt=1,\Kaot=1,\Kdot=2,\Kato=2,$ and f$\Kdto=1/2$.  Similar results hold for $B_{12}$, and $B_2$. }\label{Figure: var Da}
\end{figure}

From these results, it is evident that our ERC equations  accurately characterize $\overline{\mathbf{B}}$ and the sensogram reading (\ref{senso}) not only for small $\Da$, but for moderate $\Da$ as well.  Motivated by \citep{edwards2002testing}, we ran a series of simulations for different values of $\Da$, ranging from $\Da\approx 0.02$ to $\Da=150.$  We measured the maximum 
absolute error for each value of $\Da$, and created the curves shown in Figure \ref{Figure: var Da}. The error starts off small as expected, and increases at rates which compares favorably with our $O(\Da^2)$ prediction, and finally reaches an asymptote corresponding to roughly two percent absolute error.  Thus, although our ERC approximations (\ref{ml erc group}) and (\ref{ml erc group: diss}) are formally valid for only small values of $\Da$, their solutions agree with our finite difference approximation for moderate and large values of $\Da$.

\section{Conclusions}\label{Section: Conclusions}

Scientists are attempting to determine whether the polymerase $\eta$ and PCNA complex (denoted $EL_2$ throughout) which results from DNA translesion synthesis forms through direct binding (\ref{rxn 1}), or through a catalysis-type ligand switching process (\ref{rxn 2}).  Since fluorescent labeling techniques may modify protein behavior, label-free  optical biosensor experiments are used.  Interpreting experimental data relies on a mathematical model, and modeling multiple-component biosensor experiments results in a complicated and unwieldy set of equations.  We have shown that in experimentally relevant limits this model reduces to a much simpler set of ODEs (our ERC equations), which can be used to fit rate constants using biosensor data.  In contrast with the standard well-stirred kinetics approximation, our ERC equations accurately characterize binding when mass transport effects are significant.  This renders our ERC equations a flexible tool for estimating the rate constants in (\ref{rxns}). In turn, estimates for the rate constants in (\ref{rxns}) will reveal whether the polymerase $\eta$ and PCNA complex forms via direct binding (\ref{rxn 1}), or the catalysis-type ligand switching process (\ref{rxn 2}).


Furthermore, the consideration of both direct binding (\ref{rxn 1}) and the ligand switching process (\ref{rxn 2}) has several mathematical and physical consequences.  First, due the form of (\ref{rxns}), the species are directly coupled through the kinetics equations.  This is true even in the well-stirred limit in which $\Da\to 0$, and (\ref{c ints}) reduces to $C_1(x,0,t)=C_2(x,0,t)=1$.  However, transport effects manifested in (\ref{c ints}) nonlinearly couple the reacting species.  So we see in Figure \ref{Figure: fd pics 1} that there is a more pronounced depletion region in $B_{12}$ than in either of the other two species.  Physically, this is a consequence of the fact that either $EL_1$ or $EL_2$ must be present in order for $EL_1L_2$ to form, thus the latter is affected by depletion of the former two species.  Additionally, the multiple-component reactions (\ref{rxns}) alter the form of the sensogram reading to the lumped signal (\ref{senso}), thereby complicating parameter estimation.


In addition to establishing a firm foundation for studying the inverse problem of estimating the rate constants in (\ref{rxns}), the present work also opens the door for future work on modeling and simulating multiple-component biosensor experiments.  This includes considering other physical effects like cross-diffusion, or steric hinderance; and  comparing the finite difference method described herein to the method of lines algorithm discussed in \citep{zumbrum2013extensions}.

\appendix

\section{Parameter Values} \label{Appendix: Parameter Values}

Parameter values from the literature are tabulated below.

\begin{table}[H]
\centering
\caption{Dimensional parameter ranges, taken from references \citep{de2000calculation}, \citep{BIAcoreT200}, \citep{rich2008extracting}, \citep{yarmush1996analysis}.}
{\makegapedcells
\begin{tabular}{l|c|c|c|c}
Parameter & Rich (2008) &Yarmush &Biacore T200 & Torre\\ \hline

$k_{\ra}$ $\scriptsize ( 10^{8}\ \mathrm{cm}^3/(\mathrm{mol}\cdot \mathrm{s}))$ & $10^{-4}$--$10^{-2}$ & $.5$--$5\times 10$ & $10^{-5}$--$3\times 10$ & \\ 

$k_{\rd}$ $\scriptsize ( 10^{-3}\ \mathrm{s}^{-1})$ & $1$ & $8.9$ & $10^{-2}$--$ 10^3$ & \\

$D_1$ {\small$(10^{-7}\  \mathrm{cm}^2/\mathrm{s})$}& & & & 4.0\\

$D_2$ {\small$(10^{-7}\  \mathrm{cm}^2/\mathrm{s})$}& & & & 6.88\\

$H $ $\scriptsize (1\ \mathrm{cm})$ & $.05$ & $.04$ &  & \\ 

$L $ $\scriptsize (1\ \mathrm{cm})$ & $2$ &  &  & \\ 

$W $ $\scriptsize (1\ \mathrm{cm})$ & $1.3$ &  &  & \\ 

$R_{\mathrm{T}} $ $\scriptsize (10^{-12}\ \mathrm{mol}/\mathrm{cm^2})$ & $1.11\times 10^{-1}$--$2.33\times 10^1$ &  $2.5$--4  && \\ 

$Q$ $\scriptsize (1\ \mu L/\mathrm{min})$ & $100$--$1500$&  &$1$--$100$  & \\

$V$ $\scriptsize (1\ \mathrm{cm}/\mathrm{s})$ & $.153$--$2.88$&$.36$--$.6$ &$.001$--$1.92$  & \\ 

$C_{i,\mathrm{u}}$ $(10^{-11}\ \mathrm{mol}/\mathrm{cm}^3)$ & $2.96\times 10^{-1}$--$2\times 10^1$ & & &
\end{tabular}}\label{Table: dimensional parameters}
\end{table}
\noindent The variables $ W,\ Q$, represent the dimensional  width, and flow rate; the other dimensional variables are as in Section \ref{Section: Governing Equations}.  The flow rate is related to the velocity through the formula \citep{edwards2011transport}
\begin{equation}
V=\frac{6Q}{WH}.
\end{equation}

Using the dimensional values above, we  calculated the following \textit{extremal bounds} on the dimensionless variables.
\begin{table}[H]
\caption{Dimensionless parameters.}
\centering
\begin{tabular}{c|c}
Parameter & Bound\\
\hline
$\epsilon $&$0.02$--$0.025$ \\
$\text{Re}$ & $8.00\times 10^{-5}$--$0.36$\\
$\text{Pe}$&$ 0.12$--$523.26$\\
$\mathrm{Da}$ &$9.29\times 10^{-8}$--$1.49\times10^3$\\
${}_2K_{\mathrm{a}}$ & $2.96\times 10^{-9}$--$3.38\times 10^{8}$\\
${}_2^1K_{\mathrm{a}}$ & $2.96\times 10^{-9}$--$3.38\times 10^{8}$\\
${}_1^2K_{\mathrm{a}}$ & $2\times 10^{-7}$--$5\times 10^{6}$\\
${}_iK_{\mathrm{d}}$ & $1\times 10^{-5}$--$3.38\times 10^{8}$ \\
${}_i^jK_{\mathrm{d}}$ &  $1\times 10^{-5}$--$3.38\times 10^{8}$ \\
$D_{\mathrm{r}}$ & $0.58$\\
$F_{\mathrm{r}}$ & $0.01$--$39.28$\\
\end{tabular}
  \label{Table: dimless vals}
\end{table}
\noindent Here $\epsilon=H/L$ is the aspect ratio, and $\mathrm{Re}=VH^2/(\nu L)$ is the appropriate Reynolds number associated with our system.

The authors wish to emphasize that the bounds in Table \ref{Table: dimless vals} are na{\"i}ve extremal bounds calculated by using minimum and maximum values for the dimensional parameters in Table \ref{Table: dimensional parameters}.  In particular, the values for the dimensionless rate constants in Table \ref{Table: dimless vals} are not estimates of their true values; they are minimum and maximum values calculated using combinations of \textit{extremal values} for the parameters in Table \ref{Table: dimless vals}.  A large variation in the dimensionless rate constants is highly unlikely, since this scenario corresponds to one in which one of the association rate constants is very large, and another association rate constant very small.  We would also like to note that a large variation in some of the parameters, such as the kinetic rate constants or $\Da$, would necessitate very small values for either or both of $\Delta t$ and $\Delta x$ in our numerical method.

Furthermore, one may be concerned about the upper bound on the Reynolds number, the lower bound on the P{\'e}clet number,  and the upper bound on the Damk{\"o}hler number.  All of these extremal bounds were calculated using a flow rate of $1\ \mu \mathrm{L}/\mathrm{min}$--the slowest flow rate possible on the BIAcore T200 \citep{BIAcoreT200}.  Even with the fastest reactions, one can still design experiments to minimize transport effects by increasing the flow rate $Q$ (thus the velocity), decreasing the initial empty receptor concentration $R_{\mathrm{T}}$, and decreasing the ligand inflow concentrations $C_{1,\mathrm{u}}$ and $C_{2,\mathrm{u}}$.  In the case of the fastest reaction ${}_1k_{\ra}=3\times10^9\ \mathrm{cm}^3/(\mathrm{mol}\cdot \mathrm{s})$, we can take:
$
Q=390\ \mu\mathrm{ L}/\mathrm{min},\
V=.75\  \mathrm{cm}/\mathrm{s},\
R_{\mathrm{T}}=7.76\times 10^{-13}\ \mathrm{mol}/\mathrm{cm}^2,\
C_{1,\mathrm{u}}=C_{2,\mathrm{u}}=2.96\times10^{-12}\ \mathrm{mol}/\mathrm{cm}^3 .$
These choices yield the dimensionless parameters $\mathrm{Re}=0.09,\ \mathrm{Pe}=136.26,\ \mathrm{Da}=5.16;$ these values are perfectly in line with our analysis, and the validity of our ERC equations.

\bibliographystyle{plainnat}
\bibliography{dissertation_ref}

\end{document}